\documentstyle[12pt]{article}
\title{Group classification of heat conductivity equations
with a nonlinear source} 
\author{R.Z.~Zhdanov \\ \small Institute of Mathematics, 
3 Tereshchenkivska Street,
252004 Kyiv, Ukraine\thanks{e-mail: renat@imath.kiev.ua}
\and V.I.~Lahno \\ \small Pedagogical Institute,
2 Ostrogradskogo Street, 314000 Poltava, Ukraine
\thanks{e-mail: lahno@pdpi.poltava.ua}}
\date{}
\def\be{\begin{equation}}
\def\ee{\end{equation}}
\let\p\partial
\let\om\omega
\makeatletter\@addtoreset{equation}{section}

\newtheorem{tv}{Lemma}
\newtheorem{theo}{Theorem}

\begin{document}
\maketitle

\begin{abstract}

We suggest a systematic procedure for classifying partial differential
equations invariant with respect to low dimensional Lie algebras. This
procedure is a proper synthesis of the infinitesimal Lie's method,
technique of equivalence transformations and theory of classification of
abstract low dimensional Lie algebras. As an application, we consider
the problem of classifying heat conductivity equations in one variable
with nonlinear convection and source terms. We have derived a complete
classification of nonlinear equations of this type admitting nontrivial
symmetry. It is shown that there are three, seven, twenty eight and
twelve inequivalent classes of partial differential equations of the
considered type that are invariant under the one-, two-, three- and
four-dimensional Lie algebras, correspondingly. Furthermore, we prove
that any partial differential equation belonging to the class under
study and admitting symmetry group of the dimension higher than four is
locally equivalent to a linear equation. This classification is compared
to existing group classifications of nonlinear heat conductivity
equations and one of the conclusions is that all of them can be obtained
within the framework of our approach. Furthermore, a number of new
invariant equations are constructed which have rich symmetry properties
and, therefore, may be used for mathematical modeling of, say, nonlinear
heat transfer processes.

\end{abstract}

\section{Introduction}
\setcounter{section}{1}
\setcounter{equation}{0}

Traditionally group-theoretical, symmetry analysis of differential
equations consists of two interrelated problems. The first one is
finding the maximal Lie transformation (symmetry) group admitted by a
given equation. The second problem is one of classifying differential
equations that admit a prescribed symmetry group $G$. The principal tool
for handling both problems is the classical infinitesimal routine
developed by Sophus Lie (see, e.g., \cite{o:}--\cite{fs1:}). It reduces
the problem to finding the corresponding Lie symmetry algebra of
infinitesimal operators whose coefficients are found as solutions of
some over-determined system of linear partial differential equations
(PDEs).

Solving a classification problem for some group $G$ provides us with an
exhaustive description of differential equations that are invariant with
respect to this group and, consequently, could be analyzed by means of
the powerful Lie group technique. And it is not just a matter of
curiosity but the fundamental result that is used intensively in
applications. An experimentalist, which believes that the nature is
governed by symmetry laws, is provided with a criteria (symmetry
selection principle) for choosing a proper nonlinear model describing a
real process under investigation. Normally, a researcher has some
freedom in choosing nonlinearities of the model and it would be only
natural to take those nonlinearities that provide the highest symmetry
for the model. The classical example is the Lorentz-Poincar\'e-Einstein
relativity principle, which is to be respected by a physically
meaningful model of relativistic field theory. From the point of view of
the group theory the above principle is a requirement for a model under
study to be invariant under the Poincar\'e group (for more details, see,
e.g., \cite{fs1:,zhd97}). Consequently, finding all possible
Poincar\'e-invariant equations yields a complete account of all possible
ways to model processes of relativistic field theory by partial
differential equations.

In the overwhelming majority of papers devoted to solving classification
problems a representation of symmetry group $G$ (symmetry algebra $g$)
is fixed. Given this condition, the problem is solved by a
straightforward application of the Lie's algorithm. However, it becomes
much more complicated if no specific representation of the symmetry
algebra $g$ is given. Then utilizing the Lie's algorithm directly one
comes to the major difficulties arising from the necessity to find
maximal symmetry algebra and solve classification problem
simultaneously. A principal idea enabling to overcome the above
difficulties was suggested by Sophus Lie. Indeed, his way for obtaining
all ordinary differential equations in one variable admitting
non-trivial symmetry algebras \cite{lie24,lie27} teaches us what is to
be done in the case in question. We should first construct all the
possible inequivalent realizations of symmetry algebras within some
class of Lie vector fields. If we will succeed in doing this, then
symmetry algebras will be specified, so that we can apply directly the
Lie's infinitesimal algorithm thus getting inequivalent classes of
invariant equations. On this way, Sophus Lie has obtained his famous
classification of realizations of all inequivalent complex Lie algebras
on plane \cite{lie24,lie27}. Recently, Lie's classification has been
used by Olver and Heredero \cite{her} in order to obtain a
classification of nonlinear wave equations in (1+1) dimensions that
admit non-trivial spatial symmetries (i.e. symmetries not changing the
temporal variable). What is more, Gonzalez-Lopez, Kamran and Olver
\cite{gko91,gko94a} have classified quasi-exactly solvable models on
plane making use of their classification of real Lie algebras on plane
\cite{lie24,lie27}.

A systematic implementation of these ideas for PDEs has been suggested
by Ovsjannikov \cite{o:}. His approach is based on the concept of 
equivalence group, which is the Lie transformation group acting in the
properly extended space of independent variables, functions and their
derivatives and preserving the class of PDEs under study. It is possible
to modify the Lie's algorithm in order to make it applicable for
computing this group \cite{o:}. At the second step, the optimal system
of subgroups of the equivalence group is constructed. The next step is
utilizing the Lie's algorithm for obtaining specific PDEs belonging to
the class under study and invariant with respect to the above mentioned
subgroups.

A further development of the Ovsjannikov's approach has been undertaken
by Akhatov, Gazizov and Ibragimov \cite{ah:,ah1:}. They have obtained a
number of classification results for nonlinear gas dynamics and
diffusion equations. These ideas have been also utilized by Torrisi,
Valenti and Tracina in order to perform preliminary group classification
of some nonlinear diffusion and heat conductivity equations
\cite{tor,tor1}. Ibragimov and Torrisi have obtained a number of
important results on group classification of nonlinear detonation
equations \cite{ibr} and nonlinear hyperbolic type equations
\cite{ibr1}. Note that there are number of papers (see, e.g.,
\cite{king} and the references therein) devoted to a direct computation
of equivalence groups of some PDEs. Being somewhat more involved this
approach has a merit of giving a possibility to find {\em discrete}
equivalence groups or even non-local ones.

The Ovsjannikov's approach works smoothly provided an equivalence group
is finite-dimensional. However, if the class of PDEs under study
contains arbitrary functions of several arguments, then it could well be
that its equivalence group is infinite-parameter. The problem of
subgroup classification of infinite-parameter Lie groups is completely
open by now which makes problematic a direct application of the
Ovsjannikov's approach. Consequently, there is an evident need for
the latter to be modified to become applicable to the case of
infinite-parameter equivalence groups.

A possible way of modifying the Ovsjannikov's approach is suggested by
the manner in which physicists construct nonlinear generalizations of
the linear wave equations. They take a specific representation of the
Poincar\'e group realized on the solution set of the linear model and
require that its nonlinear generalization should inherit this symmetry
(for further details see, e.g., \cite{fs1:}). This approach makes the
classification problem fairly easy to implement, since a representation
of the symmetry algebra is fixed. A logical step forward is not to fix
{\em a priori} a specific realization of the symmetry algebra but to fix
the class of Lie vector fields within which this realization is searched
for. It is namely this idea that enabled finding principally new
nonlinear realizations of the Euclid \cite{zhd97}, Galilei
\cite{zhd97,rid1:,zhf:}, extended Galilei \cite{rid1:,zhf:},
Schr\"odinger \cite{rid1:,zhf:}, Poincar\'e \cite{zhd97} and extended
Poincar\'e \cite{rid:,lh24:} algebras. These results, in their turn,
yield broad classes of Galilei- and Poincar\'e-invariant nonlinear wave
equations.

What we suggest in the present paper is a proper combination of
the above described approaches that enables a systematic treatment
of a classification problem for the case of infinite-parameter
equivalence group admitted by the class of PDEs under study. We
perform group classification for the class of parabolic type
equations describing nonlinear heat conductivity processes
\begin{equation}
\label{1.2}
u_t=u_{xx} + F(t,\, x,\, u,\, u_x),
\end{equation}
where $u=u(t,x)$ is a smooth real-valued function, $u_t=\partial
u/\partial t,\ u_x=\partial u/\partial x$ and so on, $F$ is a
sufficiently smooth real-valued function. As shown below a direct
application of the Ovsjannikov's approach is not possible since the
equivalence group admitted by the above equation is infinite-parameter.
By this very reason, a complete group classification has been obtained
for particular cases of (\ref{1.2}) only \cite{dor}--\cite{ser}.

The paper has the following structure. In the second Section we
introduce the general method and necessary definitions and notions. The
next section is devoted to computing and analyzing the equivalence group
admitted by the class of PDEs (\ref{1.2}). In Section 4 we carry out the
preliminary group classification of (\ref{1.2}), namely, we give a
complete description of locally inequivalent PDEs of the form
(\ref{1.2}) that are invariant with respect to one-, two- and
three-dimensional Lie algebras. In the fifth Section we present all
inequivalent PDEs (\ref{1.2}) admitting four-dimensional Lie algebras.
Next, for each of thus obtained equations we compute the maximal Lie
symmetry algebra thus obtaining the complete group classification of the
corresponding models. In Section 6 we complete group classification of
invariant heat conductivity equations with nonlinear source and show
that there are no essentially nonlinear PDEs (\ref{1.2}) that admit
symmetry algebras of the dimension higher than four. The seventh Section
is devoted to an analysis of the connection of the results obtained in
the paper to other classification results for (\ref{1.2}) known to us.
It is shown that all of them can be derived from our classification of
invariant PDEs (\ref{1.2}). 

\section{Description of the method.}
\setcounter{section}{2}
\setcounter{equation}{0}

Our approach to group classification of PDEs is based on the following
facts:
\begin{itemize}
\item{PDE having a nontrivial symmetry admits some finite or infinite
dimensional Lie algebra of infinitesimal operators whose type is completely
determined by the structure constants. Furthermore, if the symmetry
algebra is infinite dimensional, then it contains as a rule some
finite dimensional Lie algebra (for example, the centerless Virasoro
algebra contains the algebra $sl(2,{\bf R})$.)}
\item{Abstract Lie algebras of the dimension up to five have been already
classified \cite{mub:,mub1:,tur:}.}
\item{Equivalence transformations preserving a class of PDEs
under study do not change the structure constants of the Lie
algebra admitted.}
\end{itemize}

Taking into account the above facts we formulate the following approach
to group classification of nonlinear heat conductivity equations
(\ref{1.2}):
\begin{enumerate}

\item[{I.}] First of all we find the most general form of infinitesimal
operators admitted by PDEs (\ref{1.2}). To this end we solve those
determining equations that do not involve the function $F$. This
yields a class ${\cal I}$ to which any symmetry of (\ref{1.2}) should
belong. Next using infinitesimal or direct approach we construct the
equivalence group $G_{\cal E}$ of the class of PDEs (\ref{1.2}).
Evidently, the group $G_{\cal E}$ sets an equivalence relation on ${\cal
I}$ (two elements of $G_{\cal E}$ are called equivalent if they are
transformed one into another with a transformation from $G_{\cal E}$).
We denote this relation as ${\cal E}$.

\item[{II.}] At the second step, we find realizations of one-, two-,
three-, four- and five-dimensional Lie algebras within the class ${\cal
I}$ up to the equivalence relation ${\cal E}$. To this end we use
the classification of low dimensional abstract Lie algebras
obtained by Mubarakzyanov \cite{mub:,mub1:}

\item[{III.}] Next, considering the obtained realizations of low
dimensional Lie algebras as symmetry algebras of PDE (\ref{1.2}) we
classify all possible forms of functions $F$ that provide invariance of
the corresponding PDE with respect to this algebra. As a result, we get a
complete classification of PDEs (\ref{1.2}) admitting Lie symmetry
algebras of the dimension up to five.

\item[{IV.}] At the last step, we apply the Lie's infinitesimal
algorithm for obtaining the maximal symmetry algebras admitted by those
PDEs (\ref{1.2}) that are invariant with respect to four- and
five-dimensional Lie algebras. This is being done straightforwardly,
since the corresponding invariant PDEs (\ref{1.2}) contains no arbitrary
functions.

\end{enumerate}

Note that the above approach does not allow for a complete group
classification of PDEs (\ref{1.2}), since there might exist realizations
of higher symmetry algebras that does not contain four- or
five-dimensional subalgebras. In fact, to get a full solution of
classification problem one still has to be able to perform an exhaustive
description of all inequivalent subalgebras of the Lie algebra of the
infinite-parameter equivalence group $G_{\cal E}$. However, in the case
under consideration our approach enables solving the group
classification problem for (\ref{1.2}) in a full generality, since there
are no essentially nonlinear PDEs of the form (\ref{1.2}) whose symmetry
algebra has a dimension higher than 4.

It is also clear, how to modify the above approach in order
to classify PDEs admitting some prescribed symmetry algebra
(say, the Galilei algebra). At the second step, one has to fix
the corresponding structure constants and find all inequivalent
realizations of the Galilei algebra within the class ${\cal I}$.
Next, the maximal symmetry algebra is computed which yields
the complete classification of Galilei-invariant PDEs of the form
(\ref{1.2}).

\section{General analysis of symmetry properties of
equation (\ref{1.2})}
\setcounter{section}{3}
\setcounter{equation}{0}

As a first step of group classification of PDE (\ref{1.2}), we find the
most general form of the infinitesimal operator of the Lie
transformation group admitted. Furthermore, we will construct the
equivalence group of the class of PDEs (\ref{1.2}).

Following the general Lie's algorithm \cite{o:,ol1:} we are looking for
an infinitesimal operator of the maximal symmetry group admitted
by (\ref{1.2}) in the form
\begin{equation} \label{3.1}
Q = \tau \p_t +\xi \p_x +\eta \p_u,
\ee
where $\tau = \tau(t,x,u), \ \xi = \xi(t,x,u), \ \eta(t,x,u)$ are
real-valued smooth functions defined in the space $X \otimes U$ of
independent $t, x$ and dependent $u = u(t,x)$ variables. The criterion
for equation (\ref{1.2}) to be invariant with respect to operator $Q$
(\ref{3.1}) reads as
\begin{equation}
\label{3.2}
\left.(\varphi^t - \varphi^{xx} -\tau F_t -\xi F_x-\eta F_u -\varphi^x
F_{u_x}\phantom{\Biggl(} )\right|_{(\ref{1.2})} =0.
\ee
Here
\begin{eqnarray} \label{3.3}
\varphi^t &=& D_t(\eta) -u_t D_t (\tau) - u_x D_t(\xi),\nonumber \\
\varphi^x &=& D_x(\eta) -u_t D_x (\tau) - u_x D_x(\xi),\\
\varphi^{xx} &=& D_x(\varphi^x) -u_{tx} D_x (\tau) - u_{xx}
D_x(\xi),\nonumber
\end{eqnarray}
$D_t, D_x$ are total differentiation operators defined in an
appropriately prolonged space $X \otimes U$:
\begin{eqnarray} \label{3.4}
D_t &=& \p_t +u_t \p_u +u_{tt} \p_{u_t} +u_{tx} \p_{u_x} +\ldots, \\
D_x &=& \p_x +u_x \p_u +u_{xx} \p_{u_x} +u_{tx} \p_{u_t} +\ldots.
\nonumber
\end{eqnarray}

Splitting (\ref{3.2}) in a usual way and solving equations that do
not involve $F$, we get the forms of the coefficients $\tau, \xi$
of the operator $Q$
\[
\tau = 2 a(t),\quad \xi = \dot a(t) x + b(t),
\]
where $a(t),\ b(t)$ are arbitrary smooth functions and
$\dot a (t) =\frac{da}{dt}$. Furthermore, the functions
$a(t),\ b(t),\ \eta = f(t,x,u)$ and $F(t,x,u,u_x)$ have
to satisfy PDE
\begin{eqnarray} \label{3.6}
&& f_t -u_x( \ddot a x +\dot b) +(f_u - 2 \dot a) F = f_{xx} +2 u_x
f_{xu} +u^2_x f_{uu} +2a F_t + \\
&& + (\dot ax+b) F_x +fF_u +f_x F_{u_x}+u_x(f_u-\dot a) F_{u_x}.
\nonumber
\end{eqnarray}

Consequently, the maximal symmetry group admitted by
equation (\ref{1.2}) is generated by an infinitesimal
operator of the form
\be \label{3.5}
Q = 2a(t) \p_t +(\dot a(t) x +b(t)) \p_x +f(t,x,u)
\p_u,
\ee
functions $a,\ b,\ f,\ F$ fulfilling the relation (\ref{3.6}).

Evidently, if we impose no restrictions on the choice
of the function $F$, then the infinitesimal operator $Q$ equals to
zero and, consequently, the symmetry group of the nonlinear
heat conductivity equation (\ref{1.2}) reduces to a trivial
group of the identity transformations. Non-trivial symmetry
groups appears, if we specify in an appropriate way
the source $F$.

As we have mentioned in Introduction, there are different ways for
constructing the equivalence group $G_{\cal E}$ for the class of PDEs
(\ref{1.2}). We use the direct method for finding the group $G_{\cal
E}$.

Let
\be \label{3.12}
\tau = \alpha(t, x, u), \ \ \xi = \beta(t, x, u), \ \ v = \gamma
(t, x, u) \ \ee
be an invertible change of variables that transforms the class
of PDEs (\ref{1.2}) into itself
\be \label{3.13}
v_\tau = v_{\xi \xi} +G(\tau, \xi, v, v_\xi).
\ee
Computing the derivative $u_x$ yields
$$
u_x = \frac{v_\tau \alpha_x +v_\xi \beta_x -\gamma_x}{\gamma_u
-v_\tau \alpha_u -v_\xi \beta_u}.
$$
On the other hand, in view of arbitrariness of the function $F$ it
follows from (\ref{3.13}) that the relation of the form
$$
u_x = g(\tau, \xi, v, v_\xi)
$$
holds. Hence we conclude that in (\ref{3.12})\ $\alpha_x = \alpha_u =0$,\
or\ $\alpha = \alpha(t), \ \dot \alpha \equiv \frac{d\alpha}{d t}\not =0$.

Computing the derivatives $u_t$, $u_{xx}$ with account of the relations
\ $\alpha_x = \alpha_u =0$\ $\Leftrightarrow$\ $\alpha = \alpha(t), \
\dot \alpha \not =0$ we get
\begin{eqnarray*}
u_t &=& v_\tau \dot \alpha(\gamma_u -v _\xi \beta_u)^{-1}
+\theta_1(\tau, \xi, v, v_\xi), \\
u_{xx}&=& v_{\xi \xi} \{ \beta^2_x(\gamma_u -v_\xi \beta_u)^{-1} +2
\beta_x \beta_u(v_\xi\beta_x -\gamma_x)(\gamma_u -v_\xi
\beta_u)^{-2}+ \\
&& +\beta^2_u(v_\xi \beta_x -\gamma_x)^2(\gamma_u-v_\xi
\beta_u)^{-3}\}+ \theta_2(\tau, \xi, v, v_\xi)
\end{eqnarray*}
with some function $\theta_2$. Taking into consideration (\ref{3.13})
yields the relation
$$
\dot \alpha (\gamma_u -v_\xi \beta_u)^2=\beta^2_x(\gamma_u-v_\xi
\beta_u)^2 +2 \beta_x \beta_u(v_\xi \beta_x -\gamma_x)(\gamma_u
-v_\xi \beta_u) +\beta^2_u (v_\xi \beta_x-\gamma_x)^2.
$$
As $\alpha, \gamma, \beta$ do not depend on $u_x$, we can
split the left-hand side of the above equation by $v_\xi$ thus
getting the system of determining equations for the functions $\alpha,
\beta, \gamma$
\begin{eqnarray*}
&& (\dot \alpha -\beta^2_x)\gamma^2_u = \gamma_x \beta_u (\gamma_x
\beta_u-2 \beta_x \gamma_u), \\
&& -2(\dot \alpha -\beta^2_x) \gamma_u \beta_u = 2 \beta^2_x \gamma_u
\beta_u, \\
&& \dot \alpha \beta^2_u=0.
\end{eqnarray*}

As $\dot \alpha \not =0$, it follows from the last equation that
$\beta_u =0$. In view of this fact system in question reduces
to a single equation
$$
(\dot \alpha -\beta^2_x) \gamma^2_u =0.
$$
Since transformation of variables (\ref{3.12}) is invertible,\ the
relation\ $\gamma_u \not =0$\ holds. Hence we get\ $\dot \alpha =
\beta^2_x$.\ Consequently,\ $\dot \alpha >0$,\ $\beta =
\pm \sqrt{\dot \alpha} x + \rho(t)$. Summing up, we conclude that the
equivalence group $G_{\cal E}$ of the class of PDEs (\ref{1.2}) reads as
\be
\label{3.14}
\bar t = T(t),\quad \bar x = \varepsilon \sqrt{\dot T(t)} x +X(t),\quad
\bar u = U(t,x,u),
\ee
where\ $\dot T(t)>0,\ U_u \not =0,\ \dot T = \frac{dT}{dt},\
\varepsilon=\pm1$.

Note that the infinitesimal method for finding the infinitesimal
operator of the equivalence group yields the following class
of operators (we skip the derivation of this formula):
\begin{eqnarray}
E&=& \alpha(t) \p_t +\Bigl [\frac{1}{2} \dot \alpha(t) x +\rho(t) \Bigr]
\p_x + \eta(t, x, u) \p_{u} +[\eta_t -\eta_{xx}\nonumber\\
&&+(\eta_u +\dot \alpha(t)) F - u_x(\frac{1}{2} \ddot \alpha(t) x +\dot
\rho(t))- 2 u_x \eta_{xu} -u^2_x \eta_{uu}] \p_F, \label{3.11}
\end{eqnarray}
where $\alpha, \rho, \eta=\eta(t,x,u)$ are arbitrary smooth functions.

It is not difficult to become convinced of the fact that 
transformations (\ref{3.14}) can be obtained from the group
transformations generated by operator (\ref{3.11}) under condition that
the latter is complemented by the discrete transformation $x \to -x$.
Consequently, both the direct and infinitesimal approaches give the same
equivalence group for the class of nonlinear heat conductivity equations
(\ref{1.2}).

\section{Preliminary group classification of equati\-on (\ref{1.2})}
\setcounter{section}{4}
\setcounter{equation}{0}

In this section we classify equations of the form (\ref{1.2}) that
admit invariance algebras of the dimension up to three. We start from
describing equations admitting one-dimensional Lie algebras, then
proceed to investigation of the ones invariant with respect to
two-dimensional algebras. Using these results we describe PDEs
(\ref{1.2}) which admit three-dimensional Lie algebras. An intermediate
problem which is being solved, while classifying invariant equations
of the form (\ref{1.2}), is describing all possible realizations
of one-, two- and three-dimensional Lie algebras by operators
(\ref{3.5}) within the equivalence relation (\ref{3.14}). One more
important remark is that PDEs that are equivalent to linear ones
are excluded from further considerations.

\subsection{Nonlinear heat equations invariant under
one-\-di\-men\-si\-on\-al Lie algebras}

All inequivalent realizations of one-dimensional Lie algebras
having the basis elements of the form (\ref{3.5}) are given by the theorem
below.
\begin{tv}
There are diffeomorphisms (\ref{3.14}) that reduce operator (\ref{3.5})
to one of the following operators:
\begin{eqnarray}
&& Q=\pm \p_t, \label{4.1} \\
&& Q = \p_x, \label{4.2} \\
&& Q = \p_u.\label{4.3}\end{eqnarray}
\end{tv}
{\it Proof.} Let an operator $Q$ have the form (\ref{3.5}). Making
the transformation (\ref{3.14}) we have
\begin{eqnarray*}
Q \to \bar Q&=& 2a\dot T\p_{\bar t}+\left[2a(\dot X +\frac{1}{2}x
\ddot T (\dot T)^{-\frac{1}{2}})\right. 
\\
&&\left. + \varepsilon(\dot a x+b) \sqrt{\dot T}\right]\p_{\bar x}
+\left[2aU_t+(\dot a x + b) U_x +fU_u\right]\p_{\bar u}.
\end{eqnarray*}

In a sequel, we have to differentiate between the cases $f =0$ and $f
\not =0$, that is why they are considered separately.
\vspace{2mm}

\noindent
{\it Case 1.}\ $f=0$. Choosing\ $U = U(u)$\ in (\ref{3.14}) yields
$$
\bar Q = 2 a \dot T \p_{\bar t}+[2 a(\dot X+\frac{1}{2}x \ddot T (\dot
T)^{-\frac{1}{2}}) +
\varepsilon(\dot a x +b)\sqrt{\dot T}]\p_{\bar x}.
$$
If\ $a =0$,\ then\ $b \not =0$ (since otherwise the operator $Q$ is
equal to zero). So that choosing as $T(t)$ in (\ref{3.14}) a solution of
the equation\ $\dot T = |b(t)|^{-2}$\ we arrive at the operator
$$
\bar Q= \pm \p_{\bar x}.
$$
Within the space reflection $x\to -x$ we may choose $Q^{'}$ in the form
$\bar Q = \p_{\bar x}$.

Given the inequality\ $a \not =0$,\ we put in (\ref{3.14})\
$\varepsilon =1$.\ Choosing as $T(t), X(t)$ solutions of system
of ordinary differential equations
\[
\dot T - \frac{1}{2|a(t)|}=0,\quad
2a(t) \dot X + b(t) \sqrt{\dot T}=0
\]
we arrive at the operator
$$
\bar Q = \pm \p_{\bar t}.
$$
\vspace{2mm}

\noindent
{\it Case 2.}\ $f \not =0$. Provided\ $a=b=0$,\ we can choose as $U$ in
(\ref{3.14}) a solution of PDE $f U_u=1$ thus getting the operator
$$
\bar Q = \p_{\bar u}.
$$
If the inequality\ $|a|+|b|\not =0$\ holds, then choosing as $U$ in
(\ref{3.14}) a solution of PDE
$$
2a U_t+(\dot a x+b) U_x + fU_u =0,\quad U_u \not=0
$$
we come to the above considered case.

It is straightforward to check that the operators (\ref{4.1}) --
(\ref{4.3}) cannot be transformed one into another with a change of
variables (\ref{3.14}). The lemma is proved. $\rhd$ 
\vspace{2mm}

Consequently, there are three inequivalent one-dimensional
Lie algebras  
$$
A^1_1=\langle\epsilon \p_t \rangle,\quad A^2_1 =\langle\p_x\rangle,
\quad A^3_1 =\langle\p_u \rangle,\quad \epsilon  =\pm 1.
$$
An easy calculation shows that the corresponding invariant equations
from the class (\ref{1.2}) have the form
\begin{eqnarray}
A^1_1&:& u_t= u_{xx}+F(x, u,u_x), \label{4.4}\\
A^2_1&:& u_t= u_{xx}+F(t, u,u_x), \label{4.5}\\
A^3_1&:& u_t= u_{xx}+F(t, x,u_x). \label{4.6}
\end{eqnarray}

To proceed further, we need the transformations from equivalence
group (\ref{3.14}) preserving the forms of the basis operators of the
above algebras. We give below the corresponding formulae
\begin{eqnarray}
A^1_1&:& \bar t = t +\lambda_1,\quad \bar x = \varepsilon x +\lambda_2,\
\ \bar u = U(x,u),
\label{4.7}\\
A^2_1&:& \bar t= t +\lambda_1,\quad \bar x = x +X(t),\quad \bar u= U(t,u),
\label{4.8}\\
A^3_1&:& \bar t = T(t),\quad \bar x = \varepsilon \sqrt{\dot T}x +X(t),
\quad \bar u= u+ U(t, x),\label{4.9}\\
&& \{\lambda_1, \lambda_2\} \subset {\bf R},\quad \varepsilon=\pm 1.\nonumber
\end{eqnarray}

\subsection{Nonlinear heat equations invariant under
two-di\-men\-si\-onal Lie algebras}

As is well-known, there are two different abstract two-dimensional Lie
algebras, namely, the commutative Lie algebra\ $A_{2.1}= \langle Q_1,
Q_2 \rangle,\ [Q_1,Q_2]=0$\ and the solvable one\ $A_{2.2}=
\langle Q_1, Q_2 \rangle ,\ [Q_1,Q_2] = Q_2$.

\begin{theo}
The list of two-dimensional Lie algebras having the basis operators
(\ref{3.5}) and defined within the equivalence relation (\ref{3.14}) is
exhausted by the following algebras:
\begin{eqnarray*}
A^1_{2.1}&=& \langle \p_t, \p_x \rangle,\quad A^2_{2.1}=
\langle \p_t,\p_u \rangle, \\
A^3_{2.1}&=& \langle \p_x, \alpha(t)\p_x+\p_u  \rangle,
\quad A^4_{2.1}= \langle \p_u, g(t,x) \p_u\rangle, g\not ={\rm const}, \\
A^5_{2.1}&=& \langle \p_x, \alpha(t) \p_x\rangle,\ \dot \alpha\equiv
\frac{d\alpha}{dt} \not =0;\\
A^1_{2.2}&=& \langle -t\p_t-\frac{1}{2}x \p_x\p_t \rangle,
\quad A^2_{2.2}= \langle -2t\p_t-x\p_x, \p_x \rangle, \\
A^3_{2.2}&=& \langle -u\p_u, \p_u \rangle,
\quad A^4_{2.2}= \langle \p_x-u \p_u,\p_u \rangle, \\
A^5_{2.2}&=& \langle \epsilon \p_t-u\p_u, \p_u \rangle, \ \ \
\epsilon=\pm 1.
\end{eqnarray*}
\end{theo}
{\it Proof.}\ Consider first the case of the commutative two-dimensional Lie
algebra. Using Lemma 1 we choose one of its basis operators (say, $Q_1$)
to be equal to one of those given in (\ref{4.1})--(\ref{4.3}). For the
sake of simplifying the form of the second basis operator $Q_2$ we make
use of equivalence transformations (\ref{4.7})--(\ref{4.9}).

If\ $Q_1=\pm \p_t$, then in view of the relation $[Q_1,\ Q_2]=0$
we obtain
$$
Q_2=\lambda \p_x+f(x,u)\p_u,\quad \lambda= \mbox{const}.
$$
Provided the equation\ $\lambda =0$\ holds, taking as $U$ in (\ref{4.7})
a solution of PDE\ $f U_u =1$\ yields the realization\ $A^2_{2.1}$.
Given the inequality\ $\lambda \not =0$\, we can choose as $U$ in
(\ref{4.7}) a solution of PDE\ $\lambda U_x+fU_u=0,\ U_u \not=0$
thus getting the realization $A^1_{2.1}$.

Let us turn now to the case when\ $Q_1=\p_x$. Then the operator
$Q_2$ takes necessarily the form
$$
Q_2 = \lambda \p_t +b(t)\p_x+f(t,u)\p_u,\quad \lambda =\mbox{const}.
$$
Provided\ $\lambda =0,\ f \not =0$, choosing as $U$ in (\ref{4.8})
a solution of PDE\ $f U_u =1$ we reduce the realization
$\langle Q_{1}, Q_{2}\rangle$ to become\ $A^3_{2.1}$.
Next, if the inequality\ $\lambda \not =0$\ holds, then
taking as $U, X$ in (\ref{4.8}) solutions of system of PDEs
$$
\lambda \dot X +b =0,\quad \lambda U_t+fU_u =0,\quad
U_u\not =0
$$
we transform the operators\ $Q_1$,\ $Q_2$\ to the basis operators
of the realization\ $A^1_{2.1}$.\ The case\ $\lambda = f =0$\ gives
rise to the realization\ $A^5_{2.1}$.

At last, consider the case when\ $Q_1= \p_u$.\ Then
$$
Q_2 = 2a(t) \p_t+(\dot ax +b)\p_x+f(t,x) \p_u.
$$
Utilizing the change of variables (\ref{4.9}), reduces the operators\
$Q_1,\ Q_2$\ to the form 
\begin{eqnarray*}
\bar Q_1&=& \p_{\bar u},\\
\bar Q_2 &=& 2a\dot T\p_{\bar t}+[2a(\varepsilon \frac{\ddot
T}{2\sqrt{\dot T}}x +\dot X) +]\varepsilon \sqrt{\dot T}
(\dot a x+b)]\p_{\bar x}\\
&& +[2aU_t+U_x(\dot a x+b) +f]\p_{\bar u}.
\end{eqnarray*}
Given the conditions\ $a=b=0$, we get the realization\ $A^4_{2.1}$
with\ $f\not = \mbox{const}$.\ If\ $a=0,\ b\not =0$, then choosing
as $T, U$ in (\ref{4.9}) solutions of system of PDEs
$$
\sqrt{\dot T}|b|=1, \quad b U_x +f =0,
$$
we get the realization\ $A^3_{2.1}$ \ $(\alpha(t)=0)$.

Provided the inequality\ $a\not = 0$\ holds, choosing
as\ $T, X, U$\ in (\ref{4.9}) solutions of system of PDEs
$$
2|a|\dot T =1, \quad 2a \dot X+\varepsilon \sqrt{\dot T}b=0,\quad
2a U_t +U_x (\dot a x +b) +f =0,
$$
transforms the operators\ $Q_1,\ Q_2$\ to become
$$
\bar Q_1= \p_{\bar u},\quad \bar Q_2 =\pm \p_{\bar t}
$$
thus yielding the realization $A^2_{2.1}$. The fact that the obtained
realizations of the two-di\-men\-si\-on\-al commutative Lie algebra are
inequivalent is established by a direct computation.

Consider now the case of the solvable two-dimensional Lie algebra.
Taking into account the results of Lemma 1 we analyze the three
possible forms of the operator $Q_2$ given in (\ref{4.1})--(\ref{4.3}).

Let us first turn to the case\ $Q_2=\pm \p_t$.\ In view of the
automorphism of the algebra under study $Q_2\to -Q_2$ we may choose
$Q_2=\p_t$. Next, using the commutation relation\ $[Q_1, Q_2] = Q_2$ \
we get
$$
Q_1 = (-t+2\lambda) \p_t +(-\frac{1}{2} x +\delta)\p_x
+f(x,u)\p_u,\quad \lambda, \delta=\mbox{const},
$$
where $f$ is an arbitrary smooth function.

Making use of the change of variables (\ref{4.7}), where\ $\lambda_1 =
-2\lambda, \ \lambda_2= -2 \delta$\ and $U$ is a solution of PDE
$$
f U_u +(\delta -\frac{1}{2} x)U_x =0,\quad U_u \not =0,
$$
we arrive at the realization\ $A^1_{2.2}$.

Consider now the case\ $Q_2=\p_x$.\ Solving the commutation relation\
$[Q_1,$\ $Q_2]=Q_2$\ yields
$$
Q_1= (-2t +2 C_1) \p_t +(-x+b(t))\p_x +f(t,u)\p_u,\quad C_1 =\mbox{const},
$$
where $b, f$ are arbitrary smooth functions.

Making the change of variables (\ref{4.8}) with\ $\lambda_1 = -C_1$\
and $X, U$ being solutions of system of PDEs
\begin{eqnarray*}
&& 2(C_1-t) \dot X+b(t)+X =0, \\
&& 2(C_1-t) U_t +f U_u=0,\quad U_u \not =0
\end{eqnarray*}
transforms the operators\ $Q_1, Q_2$ to become
$$
\bar Q_1= -2\bar t\p_{\bar t}-\bar x\p_{\bar x},\quad \bar Q_2=
\p_{\bar x},
$$
whence we get the realization\ $A^2_{2.2}$.

At last, consider the case\ $Q_2 = \p_u$.\ From the commutation relation\
$[Q_1,\ Q_2]=Q_2$\ we get the form of the operator $Q_1$
$$
Q_1= 2a(t) \p_t + (\dot a(t) x+b(t))\p_x+(-u +f(t,x))\p_u,
$$
where $a, b, f$ are arbitrary smooth functions. If\ $a=b=0$,\ then
choosing in (\ref{4.9})\ $U =-f$\ we reduce the operators\
$Q_1, Q_2$\ to become
$$
\bar Q_1= -\bar u\p_{\bar u},\quad \bar Q_2=\p_{\bar u}
$$
thus getting the realization\ $A^3_{2.2}$.

Provided\ $a=0$,\ there exists a change of variables (\ref{4.9})
reducing the operators\ $Q_1, Q_2$\ to the basis elements of the
realization\ $A^4_{2.2}$.\ The inequality\ $a \not =0$ \ gives rise
to the realization\ $A^5_{2.2}$.

The fact that the realizations obtained are inequivalent is
established by a direct verification. The theorem is proved. $\rhd$
\vspace{2mm}

Now we derive all inequivalent nonlinear heat conductivity equations
(\ref{1.2}), that admit two-dimensional Lie algebras as symmetry
algebras.

For the realizations\ $A^{1}_{2.1}$\ and\ $A^{2}_{2.1}$\ the
equations in question read as
\begin{eqnarray}
A^{1}_{2.1}&:&u_{t} = u_{xx} + \tilde F (u, u_{x}),
\label{5.1}\\
A^{2}_{2.1}&:&u_{t} = u_{xx} + \tilde F (x, u_{x}),
\label{5.2}
\end{eqnarray}
correspondingly.

Given the realization\ $A^{3}_{2.1}$\, we may use the result
of (\ref{4.5}) thus getting constraint (\ref{3.6}) for the
coefficient of the operator $Q_{2}$ in the form
$$
-\dot \alpha u_{x} = F_{u}.
$$
Hence it follows that
$$
F = - \dot \alpha uu_{x} + \tilde F (t, u_{x})
$$
with an arbitrary smooth function $\tilde F$.

So the most general PDE (\ref{1.2}) invariant with respect to the Lie
algebra\ $A^{3}_{2.1}$\ reads
\begin{eqnarray}
A^{3}_{2.1}&:& u_{t} = u_{xx} - \dot \alpha uu_{x} + \tilde
F (t, u_{x}).
\label{5.3}
\end{eqnarray}

Treating the algebra\ $A^{4}_{2.1}$\ in a similar way we represent
constraint (\ref{3.6}) as follows
$$
g_{t} = g_{xx} + g_{x}F_{u_{x}}, \quad g \not
=\mbox{const}.
$$

Given the relation\ $g_{x} = 0$,\ the function $g$ is constant, i.e.,\
$g = \mbox{const}$. This means that PDE (\ref{1.2}) becomes linear.
To avoid this we should impose the restriction\ $g_{x} \not = 0$.\ Hence,
$$
F = (g_{t} - g_{xx})g^{-1}_{x}u_{x} + \tilde F(t, x),
\quad g_{x} \not = 0.
$$

Summing up, we conclude that the class of nonlinear PDEs of the form
(\ref{1.2}) invariant with respect to the algebra\ $A^{4}_{2.1}$\ reads
as
\begin{eqnarray}
A^{4}_{2.1}&:&u_{t} = u_{xx} + (g_{t} -
g_{xx})g^{-1}_{x}u_{x} + \tilde F(t, x),\quad g_{x}
\not = 0.
\label{5.4}
\end{eqnarray}

Turn now to the algebra\ $A^{5}_{2.1}$. Inserting the
coefficients of the operator $Q_2$ into (\ref{3.6}) yields
$$
\dot \alpha u_{x} = 0,
$$
whence\ $\dot \alpha = 0.$\ This contradicts the assumption\
$\dot \alpha \not = 0$. Consequently, there are no equations of the form
(\ref{1.2}) admitting\ $A^{5}_{2.1}$\ as a symmetry algebra.

%%%part 2%%%
Treating the algebras\ $A^{i}_{2.2} \ (i = 1,\dots, 5)$\ in
a similar way we get the following invariant equations:
\begin{eqnarray}
A^{1}_{2.2}&:&u_{t} = u_{xx} + u^{2}_{x} \tilde F(u,
xu_{x});
\label{5.5}\\
A^{2}_{2.2}&:&u_{t} = u_{xx} + t^{-1} \tilde F(u,
tu^{2}_{x});
\label{5.6}\\
A^{3}_{2.2}&:&u_{t} = u_{xx} + u_{x} \tilde F(t, x);\
\label{5.7}\\
A^{4}_{2.2}&:&u_{t} = u_{xx} + u_{x} \tilde F(t,
e^{x}u_{x});
\label{5.8}\\
A^{5}_{2.2}&:&u_{t} = u_{xx} + u_{x} \tilde F(x,
e^{\epsilon t}u_{x}),\quad \epsilon =\pm 1.
\label{5.9}
\end{eqnarray}
Here $\tilde F$ is an arbitrary smooth function.

In what follows we will need equivalence transformations from the group
$G_{\cal E}$ preserving the forms of the basis operators of all
two-dimensional algebras considered above with an exception of the
algebra\ $A^{5}_{2.1}$. Omitting the derivation details we give the
the subgroups of the group $G_{\cal E}$ that do no alter
the forms of the basis operators listed in the assertion of Theorem 1.
\begin{eqnarray}
A^1_{2.1}&:& \bar t= t+\lambda_1,\ \ \bar x = x+\lambda_2,\ \
\bar u= U(u);\label{5.10}\\
A^2_{2.1}&:& \bar t = t+\lambda_1,\ \ \bar x = \varepsilon x+\lambda_2,
\ \ \bar u = u+ U(x);\label{5.11}\\
A^3_{2.1}&:& \bar t = t+\lambda_1,\ \ \bar x = x+X(t),\ \
\bar u = u+ U(t);\label{5.12}\\
A^4_{2.1}&:& \bar t = T(t),\ \ \bar x = \varepsilon \sqrt{\dot T}x+X(t),
\  \ \bar u = u+U(t,x);\label{5.13}\\
A^1_{2.2}&:& \bar t = t,\ \ \bar x = \varepsilon x,\ \
\bar u = U(u);\label{5.14}\\
A^2_{2.2}&:& \bar t = t,\ \ \bar x = x+\lambda_1 \sqrt{t},\ \
\bar u = U(u);\label{5.15}\\
A^3_{2.2}&:& \bar t = T(t),\ \ \bar x = \varepsilon \sqrt{\dot T} x+X(t),
\ \ \bar u = u;\label{5.16}\\
A^4_{2.2}&:& \bar t = t+\lambda_1,\ \ \bar x = x+X(t),\ \
u^{'}= u+e^{-x}U(t);\label{5.17}\\
A^5_{2.2}&:& \bar t = t+\lambda_1,\ \ \bar x = \varepsilon x+\lambda_2,
\ \ \bar u = u+e^{-t} U(x);\label{5.18}
\end{eqnarray}
Here\ $\{\lambda_1, \lambda_2\} \subset {\bf R},\ \varepsilon =\pm 1.$

As the above transformations do not alter the form of the basis
operators of the corresponding algebras, they can be used in order to
simplify the form of the equations admitting the latter.
An analysis shows that the only equation that can be simplified
is PDE (\ref{5.4}). 

Indeed, the change of variables (\ref{5.13}), where\ $T = t, \ X = 0$
and $U$ is an arbitrary solution of PDE
$$ 
U_{t} - U_{xx} - (g_{t} - g_{xx})g^{-1}_{x}U_{x} + \tilde
F(t, x) = 0,
$$
reduces (\ref{5.4}) to the following equation\ $(\bar t = \tau,
\bar x = \xi, \bar u = v)$:
$$
v_{\tau} = v_{\xi \xi} + (g_{\tau} - g_{\xi
\xi})g^{-1}_{\xi}v_{\xi},
$$
which is a particular case of (\ref{5.7}) (up to notations).

Thus equations (\ref{5.4}), (\ref{5.7}) are excluded from further
considerations.

\subsection{Nonlinear heat equations invariant under three-dimensional
Lie algebras}
\setcounter{subsection}{3}

We split the set of abstract three-dimensional Lie algebras into two
classes. The first class contains those algebras which are direct sums
of lower dimension ones. The remaining algebras are included into the
second class.

\subsubsection{Equations (\ref{1.2}) invariant with respect to
decomposable algebras}

The first class of Lie algebras contains two non-isomorphic algebras,
namely,\ $A_{3.1}, A_{3.2}$.\ What is more,\ $A_{3.1} = \langle Q_1, Q_2,
Q_3 \rangle$,\  $[Q_i, Q_j]=0\ (i,j =1,2,3)$,\ i.e.,\ $A_{3.1} = A_1
\oplus  A_1 \oplus  A_1=3 A_1$\ and\  $A_{3.2} = \langle Q_1, Q_2, Q_3
\rangle$,\ where\ $[Q_1, Q_2] = Q_2,\ [Q_1, Q_3] = [Q_2, Q_3] =0$,\ i.e.,\
$A_{3.2} = A_{2.2} \oplus  A_1$.

Turn first to the case of the algebra\ $A_{3.1}$.\ For describing
inequivalent realizations of this algebra we use the results of the
previous subsection on classification of inequivalent realizations of
the algebra\ $A_{2.1}$, \ namely, of the realizations,\ $A^2_{2.1}, \
A^3_{2.1}$.

Let\ $A_{2.1} = A^1_{2.1}$. Then the relations \ $Q_1 = \p_t, \ Q_2 =
\p_x$ hold, whence\ $Q_3 = f(u) \p_u$.\ Using transformation
(\ref{5.10}) we get the realization
\be 
\label{7.1} 
Q_1 = \p_t, \quad Q_2 = \p_x, \quad Q_3 =\p_u. 
\ee
Consider next the case\ $ A_{2.1} = A^2_{2.1}$.\  Then the relations\
$Q_1 = \p_t, \ Q_2 = \p_u$, \ $Q_3 = \lambda \p_x +f(x) \p_u, \ \lambda
\in {\bf R}$ hold. If\ $\lambda =0$,\  then we have the realization
\be 
\label{7.2}
Q_1 = \p_t, \quad Q_2 = \p_u, \quad Q_3 = f(x) \p_u,
\quad f' \not =0. 
\ee
If the inequality\ $\lambda \not =0$\ holds, then using (\ref{5.11})
with $U$ being a solution of PDE\ $\lambda U_x + f(x) =0$ we come
to conclusion that the operators\ $Q_i \ (i=1,2,3)$ reduce to the
form (\ref{7.1}).

Turn now to the case\ $A_{2.1} = A^3_{2.1}$. In this case we have\ $Q_1
= \p_x, \ Q_2 =\alpha(t) \p_x +\p_u$,\ whence
$$
Q_3 = 2 \lambda \p_t + b(t) \p_x +f(t) \p_u,
$$
where\ $2 \lambda \dot \alpha =0, \ \lambda \in {\bf R}$. If \ $\lambda
\not= 0$, then\ $\dot \alpha =0$. Choosing as $X, U$ in (\ref{5.12})
solutions of system of PDEs
$$ 
2 \lambda \dot X +b =0, \quad 2 \lambda U_t +f =0
$$
we reduce the operators\ $Q_i \ (i=1,2,3)$\ to the form (\ref{7.1}). 
Next, provided\ $\lambda =0$,\ the following realization is obtained
\be
\label{7.3}
Q_1 =\p_x,\quad Q_2 = \alpha(t) \p_x +\p_u,\quad Q_3 = \beta(t) \p_x
+\gamma (t) \p_u, 
\ee
where\ $\alpha(t), \beta(t), \gamma(t)$ are arbitrary smooth functions
such that the operators\ $Q_1, Q_2, Q_3$\ are linearly-independent.

Thus, within the equivalence relations defined by (\ref{3.14}), we have
the three inequivalent realizations of the algebra $A_{3.1}$, given by
formulae (\ref{7.1})--(\ref{7.3}) $Q_i \ (i=1,2,3)$. Now we proceed to
constructing the corresponding invariant equations.

Equation having as a symmetry algebra the Lie algebra (\ref{7.1})
reads as
$$
u_t = u_{xx} +G(u_x), \quad G_{u_x} \not = \mbox{const}.
$$
The restriction for $G$ guarantees that the above equation would
not be of the form (\ref{5.7}).

If the basis operators of the algebra\ $A_{3.1}$\ are given by
(\ref{7.2}), then\ $F = \tilde F(x,u_x)$,\ and the invariance condition
(\ref{3.6}) for the operator $Q_3$ reads
$$
f'' +f' {\tilde F}_{u_x} = 0, \quad f' \not =0.
$$
Hence it follows that
$$
\tilde F = -f''(f')^{-1} u_x +G(x).
$$
As established above PDE (\ref{1.2}) with\ $F = \tilde F$,\ $\tilde F$
being given by the above formula, is reduced to an equation of the form
(\ref{5.7}) and therefore is not considered in a sequel.

Next, if the basis operators of the algebra\ $A_{3.1}$\ have the form
(\ref{7.3}), then\ $F = -\dot \alpha u u_x +\tilde F(t, u_x)$\ and
what is more, the invariance condition (\ref{3.6}) for the operator\
$Q_3$\ takes the form
$$
\dot \gamma = (\dot \beta -\gamma \dot \alpha) u_x,
$$
then\ $\gamma = C_1,\ \beta =\gamma \alpha +C_2,\
\{C_1, C_2\} \subset{\bf  R}$.\ In view of this fact, we have\
$Q_3 =$ $= C_1(\alpha \p_x + \p_u) +C_2 \p_x
= C_1 Q_2 +C_2 Q_1$,\ which contradicts to the requirement of
linear independence of the operators\ $Q_i\ (i = 1, 2, 3)$.

Summing up, we conclude that there is only one realization of
the algebra\ $A_{3.1}$,\ which is a symmetry algebra of PDE
belonging to the class (\ref{1.2}) and cannot be reduced to an
equation of the form (\ref{5.7}). Namely, we have
\begin{eqnarray*}
A^{1}_{3.1} &=& \langle \p_t,\quad \p_x,\quad \p_u \rangle, \\&&
u_t = u_{xx} +G(u_x),\quad G_{u_x} \not =\mbox{const}.
\end{eqnarray*}

Let us turn now to analysis of realizations of the algebra\ $A_{3.2} =
A_{2.2} \oplus A_1$.\ In order to describe these we use the
realizations\ $A^1_{2.2},\ A^2_{2.2}$,\ $ A^4_{2.2},\ A^5_{2.2}$\ of the
two-dimensional algebra\ $A_{2.2}$ obtained in the previous subsection.

Consider first the case when\ $A_{2.2} = A^1_{2.2}$.\ Then\
$Q_1 = -t \p_t -\frac{1}{2} x \p_x,\ Q_2 = \p_t,\ Q_3 =f(u) \p_u,\
f \not =0$. It is not difficult to check that transformation
(\ref{5.14}), where\ $U$\ is a solution of PDE\ $f U_u =1$,\
reduces this triplet of operators to the form
\be
\label{7.4}
Q_1 = -t \p_t -\frac{1}{2} x \p_x, \quad Q_2 = \p_t,\quad
Q_3 = \p_u.
\ee

Next we turn to the case when\ $A_{2.2} = A^2_{2.2}$.\ With this choice
of\ $A_{2.2}$\ we get\ $Q_1 = -2 t\p_t -x \p_x,\ Q_2 = \p_x,$\ $Q_3 =
\lambda \sqrt{|t|} \p_x +f(u) \p_u,\ \lambda \in {\bf R}$.\ If\ $\lambda
=0$,\ then\ $f \not =0$ and we arrive at the realization
\be
\label{7.5}
Q_1 = -2t \p_t -x \p_x, \quad Q_2 = \p_x, \quad Q_3 =
\p_u.
\ee
Provided\ $f=0,\ \lambda \not =0$,\ we have the realization
\be
\label{7.6}
Q_1 = -2t \p_t -x \p_x,\quad Q_2 = \p_x,\quad Q_3 =
\sqrt{|t|} \p_x.
\ee
At last, if the inequality\ $\lambda f \not =0$,\ holds, then within
the transformation (\ref{5.15}) we obtain the realization
\be
\label{7.7}
Q_1 = -2t \p_t -x \p_x, \quad Q_2 = \p_x, \quad Q_3 =
\sqrt{|t|} \p_x +\p_u.
\ee

The case\ $A_{2.2} = A^{4}_{2.2}$\ gives rise to the realization
$$
Q_1 = \p_x -u \p_u,\quad Q_2 = \p_u,\quad Q_3 =
\lambda \p_t +b(t) \p_x +e^{-x} f(t) \p_u,\quad \lambda \in {\bf R}.
$$
If\ $\lambda =b =0$,\ then the following realization is
obtained
\be
\label{7.8}
Q_1 = \p_x -u \p_u, \quad Q_2 = \p_u, \quad Q_3 = e^{-x}
f(t) \p_u, \ \ f \not =0.
\ee
Next, given the conditions\ $ \lambda =0,\ b \not =0$,\ we
can choose in (\ref{5.17})\ $U = b^{-1} f$ \ and reduce
the initial operators to the form
\be
\label{7.9}
Q_1 = \p_x -u \p_u, \quad Q_2 = \p_u, \quad Q_3 = \alpha(t)
\p_x, \quad \dot \alpha \not =0.
\ee
If the inequality\ $\lambda b \not =0$\ holds, then we arrive
at the realization
\be
\label{7.10}
Q_1 = \p_x -u \p_u, \quad Q_2 = \p_u, \quad Q_3 = \p_t.
\ee

Consider next the case when\ $A_{2.2} = A^5_{2.2}$.\ Then
we have
$$
Q_1 = \epsilon \p_t -u \p_u,\quad Q_2 = \p_u,\quad
Q_3 = C_1 \p_t +C_2 \p_x +e^{-\epsilon t} f(x) \p_u,
$$
where\ $\{C_1, C_2\} \subset {\bf R},\ \epsilon =\pm 1$. Hence we
get within transformations (\ref{5.18}) and the choice of the basis
the following three realizations:
\begin{eqnarray}
Q_1 &=& \epsilon \p_t -u \p_u,\quad Q_2 = \p_u,\quad Q_3 =
e^{-\epsilon t} f(x) \p_u,\quad f \not =0,\ \epsilon =
\pm 1,  \label{7.11} \\
Q_1 &=& \epsilon \p_t -u \p_u,\quad Q_2 = \p_u,\quad Q_3 =
\p_x,\ \epsilon=\pm 1, \label{7.12} \\
Q_1 &=& \epsilon \p_t -u \p_u,\quad Q_2 = \p_u,\quad
Q_3 = \p_t +\lambda \p_x,\quad \lambda >0,\ \epsilon =
\pm 1. \label{7.13}
\end{eqnarray}

Evidently, the above obtained realizations of the algebra\ $A_{3.2}$\
and the realization\ $A^1_{3.1}$\ are inequivalent.

Now we choose from the set of so obtained realizations of
three-dimen\-sion\-al Lie algebras those which are subalgebras of
symmetry algebras of PDEs (\ref{1.2}) not reducible  to the form
(\ref{5.7}).

Equation invariant with respect to the algebra (\ref{7.4})
reads as
$$ u_t = u_{xx} +u^2_{x} G (\om),\quad \om = x u^2_x, \quad
G \not = \lambda \om^{-1},\quad \lambda \in {\bf R}.
$$

Similarly, we get PDE of the form (\ref{1.2}) admitting the
algebra\ $A_{3.2}$\ having the basis operators (\ref{7.5})
$$
u_t = u_{xx} +t^{-1} G(\om),\quad \om = t u^2_x, \quad G
\not = \lambda \sqrt{\om},\quad \lambda \in {\bf R}.
$$

If we have realization (\ref{7.6}), then\ $F = t^{-1} \tilde
F(u,tu^2_x)$\ in (\ref{1.2}). That is why invariance condition
(\ref{3.6}) for the operator $Q_3$ takes the form
$$
\epsilon\frac{1}{2 \sqrt{|t|}} u_x = 0,\quad \epsilon = \pm 1.
$$
Hence we conclude that there are no PDEs of the form (\ref{1.2})
invariant with respect to the algebra under consideration.

Provided we have realization (\ref{7.7}), the function $F$ takes the
form\ $F = t^{-1} \tilde F(u,tu^2_x)$\ and invariance condition
(\ref{3.6}) for the operator $Q_{3}$ reads as
$$
-\frac{\epsilon}{2 \sqrt{|t|}} u_x = t^{-1} \tilde F_u,
$$
where\ $\epsilon =1 \mbox{\ under \ } t>0$\ and\ $\epsilon =-1 \mbox{\
under \ } t<0$.\ Consequently,
$$
\tilde F = -\frac{1}{2} \sqrt{|\om|} u +G(\om), \quad \om = t
u^2_x
$$
and the invariant PDE is given by the following formula:
$$
u_t = u_{xx} -\frac{1}{2} t^{-1} u \sqrt{|\om|} +t^{-1} G(\om),
\quad \om = t u^2_x, \quad G \not = \lambda \sqrt{|\om|},\quad
\lambda \in {\bf R}. $$

Now we turn to the case when the operators\ $Q_i \ (i=1,2,3)$\ take one
of the forms (\ref{7.8})--(\ref{7.10}). If this is the case, then\ $F =
u_x \tilde F(t, e^xu_x)$\ and invariance condition (\ref{3.6}) for the
operator $Q_3$ is given by one of the corresponding formulae below
\begin{eqnarray*}
\dot f &=& f(1-\tilde F -\om \tilde F_\om), \quad \om = e^x u_x,
\\
- \dot \alpha &=& \alpha \om \tilde F_\om, \quad \om = e^x u_x,
\\
\tilde F_t&=& 0.
\end{eqnarray*}
Integrating these PDEs yields the forms of the functions $F$
in (\ref{1.2})
\begin{eqnarray*}
F&=& u_x (\dot f f^{-1} -1) +e^x G(t), \\
F&=&- \dot \alpha \alpha^{-1} u_x \ln |\om| +u_x G(t),\quad \om = e^x
u_x, \\
F&=& u_x G (\om),\quad \om = e^x u_x,\quad \dot G \not = \lambda
\om^{-1},\quad \lambda \in {\bf R}.
\end{eqnarray*}

A further analysis shows that only the second and the third expressions
for $F$ from the above list give rise to essentially new PDEs of the
form (\ref{1.2}).

At last, similar reasonings for triplets (\ref{7.11})--(\ref{7.13})
give the following expressions for the functions $F$
\begin{eqnarray*}
F&=&-( f+ f'')(f')u_x +e^{-\epsilon t} G(x), \\
F&=& u_x  G(e^{\epsilon t} u_x), \quad G \not = \mbox{const}, \\
F&=& u_x G (\om),\quad \om = (u_x)^{\lambda} e^{\epsilon (\lambda
t- x)},\quad \lambda >0,\quad G \not  = \mbox{const},
\quad \epsilon =\pm 1.
\end{eqnarray*}

Again, only the second and the third expressions for $F$ from the above
list give rise to essentially new PDEs of the form (\ref{1.2}).

We summarize the results on classification of nonlinear heat
conductivity equations (\ref{1.2}) invariant under the three-dimensional
Lie algebras belonging to the first class in Table 1, where we use the
following notations:
\begin{eqnarray*}
A^1_{3.1} &=& \langle \p_t, \p_x, \p_u \rangle; \\
A^1_{3.2} &=& \langle -t \p_t-\frac{1}{2} x\p_x, \p_t, \p_u \rangle; \\
A^2_{3.2} &=& \langle -2t \p_t-x \p_x, \p_x, \p_u \rangle; \\
A^3_{3.2} &=& \langle -2t \p_t-x \p_x, \p_x, \sqrt{|t|} \p_x+\p_u \rangle; \\
A^4_{3.2} &=& \langle \p_x-u \p_u, \p_u, \alpha(t) \p_x\rangle,\ \
\dot \alpha \not =0; \\
A^5_{3.2} &=& \langle \p_x-u \p_u, \p_u, \p_t \rangle; \\
A^6_{3.2} &=& \langle \epsilon \p_t-u \p_u, \p_u, \p_x \rangle; \\
A^7_{3.2} &=& \langle \epsilon \p_t-u \p_u, \p_u, \p_t+\lambda \p_x\rangle,
\ \ \lambda>0 \rangle,
\end{eqnarray*}
and what is more, $\epsilon =\pm 1$.

\subsubsection{Equations (\ref{1.2}) invariant with respect to
non-decomposable algebras}

Here we consider those three-dimensional real Lie algebras\
$A_3 =\langle Q_1, Q_2, Q_3 \rangle$\ that cannot be decomposed into
a direct sum of lower dimensional Lie algebras. The list of these
algebras is exhausted by the two semi-simple Lie algebras
\begin{eqnarray*}
A_{3.3} &:& [Q_1, Q_3] = -2 Q_2, \quad [Q_1, Q_2] = Q_1, \quad
[Q_2, Q_3] = Q_3; \\
A_{3.4} &:& [Q_1, Q_2] = Q_3, \quad [Q_2, Q_3] = Q_1, \quad
[Q_3, Q_1] = Q_2;
\end{eqnarray*}
nilpotent Lie algebra
$$
A_{3.5} : [Q_2, Q_3] = Q_1, \quad [Q_1, Q_2] = [Q_1, Q_3]
=0
$$
and six solvable Lie algebras (non-zero commutation relations are given
only)
\begin{eqnarray*}
A_{3.6} &:& [Q_1, Q_3] = Q_1, \quad [Q_2, Q_3] = Q_1+Q_2; \\
A_{3.7} &:& [Q_1, Q_3] = Q_1, \quad [Q_2, Q_3] = Q_2; \\
A_{3.8} &:& [Q_1, Q_3] = Q_1, \quad [Q_2, Q_3] = -Q_2; \\
A_{3.9} &:& [Q_1, Q_3] = Q_1, \quad [Q_2, Q_3] = q Q_2 \
(0<|q|<1) ;\\
A_{3.10} &:& [Q_1, Q_3] = -Q_2, \quad [Q_2, Q_3] = Q_1;\\
A_{3.11} &:& [Q_1, Q_3] = q Q_1-Q_2, \quad [Q_2, Q_3] =
Q_1+qQ_2, \ q>0.
\end{eqnarray*}

\begin{center}
{\it Table 1.}\ {\bf Equations (\ref{1.2}) admitting the algebras\
{\boldmath $A_{3.1}, A_{3.2}$}}
\end{center}
\begin{center}
\begin{tabular}{| l | c| }   \hline
& \\
Algebra & Function \ $F$ \\  [2mm] \hline
& \\
$A^1_{3.1}$&$G(u_x), \ \ G_{u_x} \not= \lambda, \ \ \lambda \in {\bf R}$ \\
 [2mm] \hline
& \\
$A^1_{3.2}$&$ u^2_x G(\om), \ \ \om = x u_x, \ \ G \not= \lambda
\om^{-2} , \ \ \lambda \in {\bf R}$ \\ [2mm] \hline
& \\
$A^2_{3.2}$&$ t^{-1} G(\om), \ \ \om = t u^2_x, \ \
G \not =\lambda \sqrt{\om},\ \ \lambda \in {\bf R}$ \\ [2mm] \hline
& \\
$A^3_{3.2}$ & $ -\frac{1}{2} t^{-1} u \sqrt{|\om|} +t^{-1} G(\om),
\ \ \om = t u^2_x $ \\ [2mm] \hline
& \\
$A^4_{3.2}$ & $- \dot \alpha \alpha^{-1} u_x \ln |\om|+u_x G(t),
 \ \ \dot \alpha  \not =0, \om = e^{x} u_x $ \\ [2mm] \hline
& \\
$A^5_{3.2} $ & $ u_x G(\om), \ \ \om = e^x u_x, \ \
G \not =\lambda \om^{-1}, \ \ \lambda \in {\bf R}$ \\ [2mm] \hline
& \\
$A^6_{3.2}$ & $u_x G(\om), \ \ \om = e^{\epsilon t} u_x, \ \
G \not = \lambda \om^{-1}, \ \ \lambda \in {\bf R}, \ \ \epsilon =\pm 1
$ \\ [2mm] \hline
& \\
$A^7_{3.2}$ & $u_x G(\om), \ \ \om = ( u_x)^\lambda
e^{\epsilon(\lambda t-x)}, \ \ \lambda >0, \ \ G \not = \mbox{const}, \ \
\epsilon =\pm 1$ \\ [2mm] \hline
\end{tabular}
\end{center}

While constructing inequivalent realizations of the above algebras
within the class of operators (\ref{3.5}), we use wherever possible
the classification results obtained for the lower dimensional Lie
algebras.

Consider first the semi-simple algebras. Let\ $A_3 = A_{3.3}$.\ Then \
$Q_1, Q_2$\ satisfy the commutation relation\ $[Q_1, Q_2] = Q_1$\ and
form a basis of a two-dimensional Lie algebra isomorphic to\ $A_{2.2}$.
Indeed, choosing\ $Q_1 = Q^{'}_2$\ $Q_2 = -Q^{'}_1$\ we see that \
$[Q^{'}_1, Q^{'}_2] = -[Q_2, Q_1] = Q_1 = Q^{'}_2$.\ Thus we can use the
results on classification of the algebra\ $A_{2.2}$.\ According to the
results of Subsection 4.2 studying realizations of the algebra\
$A_{3.3}$\ reduces to finding the form of the operator $Q_3$ for each
pair of the operators\ $Q_1, Q_2$\ given below
\begin{eqnarray}
1)&& Q_1 = \epsilon \p_t, \quad Q_2 = t \p_t +\frac{1}{2} x \p_x;
\nonumber \\
2)&& Q_1=\p_x, \quad Q_2 = 2t \p_t +x\p_x; \label{7.14}\\
3)&& Q_1=\p_u, \quad Q_2 = u \p_u -\p_x; \nonumber \\
4)&& Q_1 = \p_u, \quad Q_2 = u \p_u -\epsilon \p_t. \nonumber
\end{eqnarray}
Here $\epsilon =\pm 1$.

One more remark is that the form of the operator $Q_3$ can be simplified
with the use of transformations (\ref{5.14}), (\ref{5.15}),
(\ref{5.17}), (\ref{5.18}).

Let\ $Q_1, Q_2$\ be given by the first formula from (\ref{7.14}).
Then it follows from the commutation relations
\be
\label{7.15}
[Q_1, Q_3] = -2 Q_2, \quad [Q_2, Q_3] = Q_3
\ee
that
$$
Q_3 = -\epsilon t^2\p_t -\epsilon tx \p_x +x^2 f(u) \p_u.
$$
Given the condition\ $f(u) \not =0$,\ change of variables (\ref{5.14})
with\ $\varepsilon =1$\ and $U$ being a solution of PDE\ $f U_u =1$\
reduces the operator $Q_3$ to the form
$$
\bar Q_3 = -\epsilon \bar t^2 \p_{\bar t} -\epsilon \bar t \bar x
\p_{\bar x} + \bar x \p_{\bar u}.
$$
Consequently, we get the realization
\be
\label{7.16}
Q_1 = \p_t, \quad Q_2 = t \p_t +\frac{1}{2} x \p_x, \quad
Q_3 = -t^2 \p_t -tx \p_x + \varepsilon x^2\p_u, \quad
\varepsilon =0,1.
\ee

Let\ $Q_1, Q_2$\ be given by the second formula from (\ref{7.14}).
Checking commutation relations (\ref{7.15}) yields that there is no
operator $Q_3$ of the form (\ref{3.5}) which enables extending the
algebra\ $A_{2.2}$\ to the algebra\ $A_{3.3}$.\ The same assertion holds
for the remaining pairs of operators from (\ref{7.14}).

Thus there exists a unique realization of the algebra\ $A_{3.3}$\ that
is given by (\ref{7.16}). In this case,\ $F = u^2_x \tilde F(u, \om),\
\om =x u_x$\ and, consequently, invariance condition (\ref{3.6})
for the operator $Q_3$ takes the form
$$
\varepsilon_1 [\om^2 \tilde F_u +2 \om^2 \tilde F_\om +4 \om \tilde
F +2] =-\epsilon \om.
$$
Provided\ $\varepsilon_1 =0$,\ we get the equality\ $ \om =0$\
whence it follows that the only possible value of $\varepsilon_1$
is\ $\varepsilon_1 =1$.\ With this condition,
$$
\tilde F = - \frac{\epsilon}{4} -\om^{-1} +\om^{-2} G(2u -\om).
$$
Hence we conclude that equation (\ref{1.2}) invariant with respect to
the algebra
$$
A^1_{3.3} = \langle \epsilon \p_t,\quad t \p_t +\frac{1}{2} x \p_x,\quad
-\epsilon t^2 \p_t -\epsilon tx \p_x +x^2 \p_u \rangle
$$
reads as
$$
u_t = u_{xx} + \epsilon \frac{1}{4} u^2_x -x^{-1} u_x +x^{-2}
G(2u-xu_x),\quad \epsilon=\pm 1.
$$

On having used the equivalence transformation
$$
t \to t, \quad x \to x, \quad u \to -u,
$$
we may choose\ $\epsilon=1$.

Note that the algebra\ $A^1_{3.3}$\ is isomorphic to the Lie algebra
of pseudo-orthogonal group\ $O(1,2)$.

Turn now to the algebra\ $A_{3.4}$.\ It does not contain a two-dimensional
subalgebra and we use the classification results for one-dimensional
algebras (Subsection 4.1). According to these results the operator $Q_1$
is reduced to one of the following inequivalent forms
\be
\label{7.17}
\pm  \p_t, \quad  \p_x, \quad \p_u.
\ee
 
Given the relation\ $Q_1 =\pm  \p_t$,\ we verify that there are no
operators $Q_2, Q_3$ of the form (\ref{3.5}) satisfying together with
$Q_1$ the commutation relations
$$ [Q_1, Q_2] = Q_3, \quad [Q_2, Q_3] = Q_1,
\quad [Q_3, Q_1] = Q_2.
$$
Consequently, the class of operators (\ref{3.5}) does not contain
operators\ $Q_2, Q_3$\ that extend a realization of the
one-dimensional algebra $\langle Q_1\rangle$ to a realization of the
algebra\ $A_{3.4}$.\ The same assertion holds true for the remaining
realizations of the operator $Q_1$. Summing up we conclude that there is
no PDE of the form (\ref{1.2}) whose symmetry algebra contains a
three-dimensional algebra isomorphic to\ $A_{3.4}$.

The algebra\ $A_{3.5}$\ contains the commuting subalgebra having
the basis operators\ $Q_1, Q_2$.\ Since the latter is isomorphic
to the Lie algebra\ $A_{2.1}$,\ we can use the results of Subsection 4.2.
In view of these we conclude that there are three inequivalent realizations
of the algebra\ $A_{2.1}$\ which might be invariance algebras of equations
of the form (\ref{1.2}), namely,
\begin{eqnarray}
A^1_{2.1}&=&\langle  \p_t, \p_x\rangle ; \nonumber \\
A^2_{2.1}&=&\langle \p_t, \p_u \rangle ;\label{7.18} \\
A^3_{2.1} &=&\langle  \p_x, \alpha(t) \p_x+\p_u\rangle . \nonumber
\end{eqnarray}

Therefore, while considering the algebra\ $A_{3.5}$\ we can suppose
that\ $Q_1, Q_2$ are given by one of the formulae (\ref{7.18}). In order
to simplify the form of the operator $Q_3$ we use transformations
(\ref{5.10}), (\ref{5.11}), (\ref{5.12}), respectively.

Let the operators\ $Q_1, Q_2$\ form a basis of the algebra\
$A^1_{2.1}$. If\ $Q_1=\p_x,\ Q_2=\p_t$,\ then analyzing the commutation
relations
\be
\label{7.19}
[Q_1, Q_3] =0, \quad [Q_2, Q_3] = Q_1
\ee
yields that the class of operators (\ref{3.5}) does not contain an
operator $Q_3$ which forms together with\ $Q_1, Q_2$\ a basis of the
algebra\ $A_{3.5}$.

Next, provided\ $Q_1=\p_x,\ Q_2=\p_t$, it follows from (\ref{7.19})
that
$$
Q_3 = (t +\lambda_2) \p_x+f(u) \p_u.
$$

There is a transformation (\ref{5.9}) that reduce $Q_3$ to the form
\be
\label{7.20}
 Q_3 = t \p_x + \epsilon \p_u,\quad \epsilon =0,1.
\ee

The most general PDE (\ref{1.2}), which is invariant with respect
to the algebra\ $A^1_{2.1}$\ reads
\be
\label{7.21}
u_t = u_{xx} +\tilde F(u, u_x).
\ee
That is why, condition for PDE (\ref{1.2}) to be invariant under
the obtained realization of the algebra\ $A_{3.5}$ coincides
with (\ref{3.6})
$$
-u_x = \epsilon \tilde F_u,
$$
whence it follows that in (\ref{7.20})\ $\epsilon =1$ and
in (\ref{7.21})
$$
\tilde F = -u u_x+G(u_x).
$$
Thus the algebra\ $A^1_{3.5} = \langle \p_x, \p_t,
t\p_x+\p_u \rangle$\ is the invariance algebra of the nonlinear
PDE
$$
u_t = u_{xx} -u u_x +G(u_x).
$$

Analysis of the cases when the operators\ $Q_2, Q_2$ form bases of the
algebras\ $A^2_{2.1}, A^3_{2.1}$\ is carried out in a similar way. As a
result, we get three more realizations that are invariance algebras of
PDEs of the form (\ref{1.2})
\begin{eqnarray*}
A^2_{3.5} &=& \langle \p_u, \p_t, t \p_u +\lambda \p_x \rangle, \\
&& u_t = u_{xx} +\lambda ^{-1} x +G(u_x),\quad \lambda >0; \\
A^3_{3.5} &=& \langle \p_u, \p_x, x \p_u +b(t) \p_x \rangle, \\
&& u_t = u_{xx} -\frac{1}{2} \dot b(t) u^2_x +G(t),\quad \dot b(t) \not =0;
\\ A^4_{3.5}&=& \langle \p_u, \p_x, x \p_u +\lambda \p_t \rangle, \\
&& u_t = u_{xx} +G(\om), \ \ \om = t-\lambda u_x,\quad \lambda \not
=0; \\
A^5_{3.5}&=& \langle \p_u+2 \lambda t \p_x, \p_x, x\p_u +2 \lambda t[
t\p_t+x\p_x -u \p_u] \rangle, \\
&& u_t = u_{xx} -2 \lambda u u_x +t^{-3} G(\om),\quad \om = u_x t^2
-\frac{t}{2 \lambda},\quad \lambda \not =0.
\end{eqnarray*}

Next we consider the solvable algebras. These algebras have a common
feature, namely, they contain commutative two-dimensional subalgebras
with basis operators\ $Q_1, Q_2$.\ That is why, analysis of these
algebras is similar to that of the algebra\ $A_{3.5}$.

Consider, for example, the algebra\ $A_{3.9}$.\ Since the admissible
pairs of the operators\ $Q_1, Q_2$ are known, all what should be done is
to check the commutation relations
\be
\label{7.22}
[Q_1, Q_3] =Q_1,\quad [Q_2, Q_3] = qQ_2,\quad 0 <|q|<1,
\ee
the operator $Q_3$ being of the form (\ref{3.5}).

Let the operators\ $Q_1, Q_2$\ form a basis of the algebra\ $A^1_{2,1}$.\
If\ $Q_1 = \p_t,\ Q_2 = \p_x$,\ then it follows from commutation relations
(\ref{7.22}) that within transformations (\ref{5.10})\ $q = \frac{1}{2}$\
and furthermore
$$
Q_3 = t \p_t +\frac{1}{2}x\p_x+\epsilon u \p_u, \quad \epsilon
 =0,1.
$$
After checking the condition of invariance of equation (\ref{1.2})
under the obtained realization of the algebra\ $A_{3.5}$\ we see
that, given the relation $\epsilon =0$, the invariant PDE reads as
$$
u_t = u_{xx} +u^2_x G(u)
$$
and with $\epsilon =1$ the invariant PDE takes the form
$$
u_t = u_{xx} +G(\om),\quad \om = u u^2_x.
$$

Provided\ $Q_1 = \p_x,\ Q_2 = \p_t$,\ we get from commutation relations
(\ref{7.22}) that $q=2$. This contradicts to the condition $0<|q|<1$.

Let the operators\ $Q_1,\ Q_2$\ form a basis of the realization\
$A^2_{2.1}$. If\ $Q_1 = \p_t,\ Q_2 = \p_u$,\ then
$$
Q_3 = t \p_t +\frac{1}{2} x \p_x +qu \p_u,\quad 0<|q|<1.
$$
Provided\ $Q_1 = \p_u,\ Q_2 = \p_t$, we get the following form
of the operator $Q_3$:
$$
Q_3 = qt \p_t +\frac{1}{2} qx \p_x +u\p_u,\quad 0<|q|<1.
$$

Thus we have obtained two distinct realizations of the
algebra\ $A_{3.9}$
\begin{eqnarray*}
L_1&=& \langle \p_t, \p_u, t \p_t +\frac{1}{2} x \p_x +qu \p_u
\rangle, \quad  0<|q|<1; \\
L_2&=& \langle \p_u,\p_t, qt \p_t +\frac{1}{2} qx \p_x +u \p_u
\rangle, \quad  0<|q|<1.
\end{eqnarray*}
These two realizations can be unified in the following way:
$$
Q_1 = \p_t,\quad  Q_2 = \p_u,\quad Q_3 = t \p_t +\frac{1}{2} x\p_x
+qu \p_u,\quad q \not =0, \pm 1.
$$
The corresponding invariant equation reads
$$
u_t = u_{xx}+ x^{2(q-1)} G(\om), \quad \om = x^{1-2q} u_x.
$$

At last, let us consider the case when the operators\ $Q_1, Q_2$\
form a basis of the realization\ $A^3_{2.1}$.\ This case is
handled in the same way as the previous one and the results are as follows.
We get one more realization of the algebra\ $A_{3.9}$ whose basis
is formed by the operators
$$
Q_1 = \p_x, \quad Q_2 = \p_u +\lambda |t|^{\frac{1}{2}(1-q)} \p_x,\quad
Q_3 = 2t \p_t +x\p_x +qu \p_u,
$$
where\ $q\not=0, \pm 1;\ \lambda \in {\bf R}.$ The corresponding
invariant equation reads
$$
u_t = u_{xx} -\frac{1}{2} \lambda (1-q) |t|^{-\frac{1}{2} (1+q)} u
u_x+|t|^{\frac{1}{2} (q-1)} G(\om)
$$
with
$$
\om =|t|^{\frac{1}{2} (1-q)}u_x.
$$

The remaining solvable Lie algebras are handled in an analogous way.
The results on classification of nonlinear heat conductivity
equations (\ref{1.2}) admitting the three-dimensional Lie algebras
from the second class are summarized in Table 2, where the following
notations are used:
\begin{eqnarray}
A_{3.3}^{1} &=& \langle \partial_{t}, t \partial_{t} +
\frac{1}{2} x \partial_{x}, -t^{2} \partial_{t} -
tx \partial_{x} + x^{2} \partial_{u} \rangle ,\nonumber\\ [3mm]
A_{3.5}^{1} &=& \langle \partial_{x},  \partial_{t}, t
\partial_{x}+ \partial_{u} \rangle ,\nonumber\\ [3mm]
A^2_{3.5} &=& \langle \p_u, \p_t, t \p_u +\lambda \p_x\rangle,\quad
\lambda>0, \nonumber\\ [3mm]
A^3_{3.5} &=& \langle \p_u, \p_x, x \p_u +b(t) \p_x\rangle,\quad
\dot b\not =0, \nonumber\\ [3mm]
A^4_{3.5} &=& \langle \p_u, \p_x, x \p_u +\lambda \p_t\rangle,\quad
\lambda\not =0, \nonumber\\ [3mm]
A^5_{3.5} &=& \langle \p_u+2 \lambda t \p_x, \p_x, x \p_u +2\lambda t
[t \p_t +x \p_x-u \p_u]\rangle,\quad \lambda\not =0, \nonumber\\ [3mm]
A_{3.6}^{1} &=& \langle \partial_{u}, \p_t, \quad t
\partial_{t} + \frac{1}{2}x\partial_{x} +(u+t)\p_u \rangle ,\nonumber\\ [3mm]
A_{3.6}^{2} &=& \langle \partial_{x}, \quad \partial_{u} -\frac{1}{2}
\ln|t| \p_x, \quad 2t \p_t +x\p_x+u \p_u\rangle ,\nonumber\\ [3mm]
A_{3.6}^{3} &=& \langle \partial_{u}, \quad \partial_{x}, 2t \p_t
+x\p_x +(u+x)\p_u \rangle;\nonumber\\ [3mm]
A^4_{3.6} &=& \langle \p_u, \alpha \p_x, \alpha^2 (\dot \alpha)^{-1}
\p_t+(1+\alpha) x \p_x +[(1-\alpha) u +x] \p_u \rangle ,\quad
\alpha = \alpha(t),\ \dot \alpha \not =0\nonumber\\
&&\mbox{and}\ \alpha^2 \ddot \alpha+2 (\dot \alpha)^2=0;\nonumber\\
A^1_{3.7}&=& \langle \p_t, \p_u, t \p_t +\frac{1}{2} x \p_x +u \p_u
\rangle ; \nonumber\\ [3mm]
A_{3.7}^{2} &=& \langle \partial_{x}, \quad \p_u, 2 t \p_t +x \p_x +u
\p_u \rangle ,\nonumber\\ [3mm]
A_{3.8}^{1} &=& \langle \partial_{t}, \quad  \partial_{u}, \quad t
\partial_{t} + \frac{1}{2} x \partial_{x}-u \p_u \rangle ,\nonumber\\ [3mm]
A_{3.8}^{2} &=& \langle \partial_{x}, \quad  \partial_{u} +
\lambda t \p_x, 2t \p_t +x \p_x -u \p_u\rangle,\quad \lambda \in {\bf R};
\nonumber\\ [3mm]
A_{3.9}^{1} &=& \langle \partial_{t}, \quad \partial_{x}, \quad t
\partial_{t} + \frac{1}{2} x \partial_{x} \rangle ,\nonumber\\ [3mm]
A_{3.9}^{2} &=& \langle \partial_{t},\quad  \partial_{x}, \ \
  t \partial_{t} + \frac{1}{2} x \partial_{x}+u \p_u
\rangle ;\nonumber\\ [3mm]
A_{3.9}^{3} &=& \langle \partial_{t}, \quad  \partial_{u}, \ \
t \partial_{t} + \frac{1}{2} x \partial_{x}+qu \p_u\rangle,\quad
q\not =0, \pm 1; \nonumber\\ [3mm]
A_{3.9}^{4} &=& \langle \partial_{x}, \quad \partial_{u} + \lambda
|t|^{\frac{1}{2}(1-q)} \partial_{x}, \quad 2t \partial_{t} + x
\partial_{x} +qu \p_u\rangle,\quad  0<|q|<1,\ \lambda \in {\bf R};
\nonumber\\ [3mm]
A_{3.10}^{1} &=& \langle \partial_{x}, \quad \lambda t
\partial_{x} + \partial_{u},\quad  -\lambda (t^{2} + \lambda
^{-2}) \partial_{t} - \lambda tx \partial_{x} + (\lambda tu
- x) \partial_{u}\rangle,\quad \lambda \not = 0;\nonumber\\ [3mm]
A^1_{3.11}&=&\langle \p_x, \quad \alpha \p_x +\p_u, \quad -(\dot
\alpha)^{-1} (1+\alpha^2) \p_t +(q-\alpha)
x \p_x +[(\alpha+q)u -x] \p_u \rangle,\nonumber\\
&&q>0;\quad \alpha=\alpha(t),\ \dot \alpha\not =0\ \mbox{and}\
(1+\alpha^2) \ddot \alpha=2q(\dot \alpha)^2.\nonumber
\end{eqnarray}

Ordinary differential equations
\begin{eqnarray}
&&\alpha^2 \ddot \alpha+2 (\dot \alpha)^2=0,\label{4.50}\\
&&(1+\alpha^2) \ddot \alpha=2q(\dot \alpha)^2.\label{4.51}
\end{eqnarray}
can be solved by quadratures. However their general solutions are
defined implicitly and  cannot be expressed via elementary functions.

\newpage
\begin{center}
{\it Table 2.}\ {\bf Equations (\ref{1.2}) admitting
three-dimensional Lie algebras from the second class}
\end{center}
\begin{center}
\begin{tabular}{| l | c| }   \hline
& \\
Algebra & Function \ $F$ \\  [2mm] \hline
& \\
$A^1_{3.3}$&$\frac{1}{4} u^2_x - x^{-1} u_x +x^{-2} G(\om), \hskip
5mm \om = 2u -x u_x$ \\ [2mm] \hline
& \\
$A^1_{3.5}$&$-u u_x + G(u_x)$ \\ [2mm] \hline
& \\
$A^2_{3.5}$&$\lambda^{-1} x+ G(u_x), \quad \lambda > 0, \quad
G_{u_xu_x}\ne 0$ \\
[2mm] \hline
& \\
$A^3_{3.5}$&$-\frac{1}{2} \dot b(t) u^2_x +G(t),
\quad \dot b \not =0 $ \\ [2mm] \hline
& \\
$A^4_{3.5}$&$ G(\om), \quad \om = t-\lambda u_x, \quad \lambda \not =0 \
\ G_{\omega\omega}\ne 0$ \\ [2mm] \hline
& \\
$A^5_{3.5}$&$-2 \lambda u u_x + t^{-3} G(\om), \quad \om = u_x t^2
-\frac{t}{2 \lambda}, \quad \lambda \not =0$ \\ [2mm] \hline
& \\
$A^1_{3.6}$&$2\ln|u_x|G(\om), \quad \om = x^{-1} u_x$ \\ [2mm] \hline
& \\
$A^2_{3.6}$&$\frac{1}{2} t^{-1} u u_x +|t|^{-\frac{1}{2}}G(u_x)$, \\
[2mm] \hline
& \\
$A^3_{3.6}$&$|t|^{-\frac{1}{2}} G(\om), \hskip
5mm \om = t^{-1}  u^2_x,\quad G\ne {\rm const}, \sqrt{\omega}$
\\ [2mm] \hline
& \\
$A^4_{3.6}$&$-\dot \alpha u u_x +\alpha^{-6} \exp(2 \alpha^{-1})
G(\om), \quad \om = u_x\alpha^4-\frac{2}{3} \alpha^3$ \\ [2mm] \hline
& \\
$A^1_{3.7}$&$ G(\om), \hskip
5mm \om = x^{-1} u_x,\quad G_{\omega\omega}\ne 0$ \\ [2mm]
\hline
& \\
$A^2_{3.7}$&$|t|^{-\frac{1}{2}}G(u_x),\quad G_{u_xu_x}\ne 0$
\\ [2mm] \hline
\end{tabular}
\end{center}

\begin{center}
\begin{tabular}{| l | c| }   \hline
& \\
Algebra & Function \ $F$ \\  [2mm] \hline
& \\
$A^1_{3.8}$&$ x^{-4} G(\om), \hskip
5mm \om = x^3 u_x,\quad G_{\omega\omega}\ne 0$ \\ [2mm]
\hline
& \\
$A^2_{3.8}$&$-\lambda u u_x +|t|^{-\frac{3}{2}}G(\om), \quad
\om = t u_x,\quad \lambda\in {\bf R},\quad \lambda^2
+G_{\omega\omega}\ne 0$ \\ [2mm] \hline & \\
$A^1_{3.9}$&$ u^2_x G(u),\quad G_u \ne 0$\\ [2mm] \hline
& \\
$A^2_{3.9}$&$ G(\om), \quad  \om = u^{-1} u^2_x,\quad
G_{\omega}\ne 0 $\\ [2mm] \hline
& \\
$A^3_{3.9}$&$ x^{2(q-1)} G(\om), \quad  \om = x^{1-2q} u_x,\quad
G_{\omega\omega}\ne 0$ \\ [2mm]
\hline
& \\
$A^4_{3.9}$&$ -\frac{1}{2} \lambda (1-q) |t|^{-\frac{1}{2} (1+q)} u
u_x +|t|^{\frac{1}{2} (q-2)} G(\om), \ \om = |t|^{\frac{1}{2}(1-q)}
u_x,\quad \lambda^2+G_{\omega\omega}^2\ne 0$\\ [2mm] \hline
& \\
$A^1_{3.10}$&$ -\lambda u u_x +(t^2+\lambda^{-2})^{-\frac{3}{2}}
G(\om), \ \om = \lambda u_x (t^2+\lambda^{-2}) -t,\quad \lambda\ne 0
$\\ [2mm] \hline
& \\
$A^1_{3.11}$&$ -\dot \alpha u u_x +(1+\alpha^2)^{-\frac{3}{2}} \exp(q
\arctan \alpha) G(\om), \ \om = u_x (1+\alpha^2) -\alpha $\\ [2mm] \hline
\end{tabular}
\end{center}
\vspace{2mm}

The general solution of (\ref{4.50}) reads as
$$
\int^\alpha\exp(-2\xi^{-1}) d \xi = \lambda t +\lambda_1,
\quad \{\lambda, \lambda_1\} \subset {\bf R}, \quad \lambda \not =0;
$$
and the general solution of (\ref{4.51}) is given by the formula
$$
\int^\alpha \exp(-2q \arctan\xi) d \xi = \lambda t +\lambda_1,\quad
\{\lambda, \lambda_1\}\subset {\bf R}, \quad \lambda \not =0.
$$

One more important remark is that the obtained realizations of
three-dimensional Lie algebras are inequivalent. This means, in
particular, that the corresponding invariant equations are
inequivalent as well.

%%%part 3%%%
\section{Complete group classification of equati\-ons (\ref{1.2})
invariant under four-dimensional Lie algebras}
\setcounter{section}{5}
\setcounter{equation}{0}

In this section we carry out group classification of nonlinear heat
conductivity equations (\ref{1.2}) admitting four-dimensional Lie
algebras. To this end, we use the known classification of abstract
four-dimensional Lie algebras \cite{mub:}. Furthermore for each
invariant equation we compute the maximal in Lie's sense symmetry
algebra thus completing the classification.

As calculations performed for constructing inequivalent realizations of
four-dimensional Lie algebras within the class of operators (\ref{3.5})
are essentially the same as those used when we study three-dimensional
ones, we will concentrate on giving the final results omitting
calculation details. As above, we should differentiate between the cases
of decomposable and non-de\-com\-pos\-able four-dimensional Lie
algebras.

\subsection{PDEs (\ref{1.2}) invariant under decomposable
four-dimensional Lie algebras}

The class of decomposable four-dimensional Lie algebras (regarded in a
sequel as the first class) contains twelve algebras:\ $4 A_1 =
A_{3.1}\oplus A_1,\ A_{2.2}\oplus 2A_1= A_{3.2}\oplus A_1,\ 2
A_{2.2}=A_{2.2}\oplus A_{2.2},\ A_{3.i}\oplus A_1\ (i = 3,4, \ldots
11)$. We preserve the notations of the previous section. What is more,\
$A_{3.i} = \langle Q_1, Q_2, Q_3 \rangle\ (i=1,2,\ldots, 11), \ A_{1} =
\langle Q_4 \rangle$.

An analysis shows that within the class of operators (\ref{3.5}) there
are four inequivalent realizations of the algebra\ $2A_{2.2}$ which are
invariance algebras of PDEs of the form (\ref{1.2}). We give these
realizations below together with the corresponding invariant equations.
\begin{eqnarray}
2 A^1_{2.2} &=& \langle -t \p_t -\frac{1}{2} x \p_x, \p_t, \p_u,
e^u\p_u \rangle,\nonumber\\
&& u_t = u_{xx} -u^2_x +\frac{\lambda}{x} u_x, \quad \lambda \in {\bf R};
\label{5.1*} \\
2 A^2_{2.2} &=& \langle -2t \p_t - x \p_x, \p_x, \p_u,
e^u\p_u \rangle,\nonumber\\
&& u_t = u_{xx} -u^2_x +\lambda \frac{u_x}{\sqrt{|t|}},
\quad \lambda \in {\bf R}; \label{5.2*} \\
2 A^3_{2.2} &=& \langle -2t \p_t - x \p_x, \p_x, -u\p_u+\lambda
\sqrt{|t|} \p_x, \p_u \rangle,\nonumber\\
&& u_t = u_{xx} +\frac{\lambda \epsilon u_x}{4\sqrt{|t|}}
\ln|t u^2_x| +\frac{\beta u_x}{\sqrt{|t|}}; \label{5.3*} \\
&&  \epsilon =1 \mbox{\ for \ } t>0
\mbox{ \ and \ }\epsilon=-1 \mbox{\ for \ } t<0, \ \lambda \not =0, \ \
\beta \in {\bf R}; \nonumber \\
2 A^4_{2.2} &=& \langle  \p_x -u\p_u, \p_u, \frac{1}{\lambda}\p_t,
e^{\lambda t}\p_x \rangle,\nonumber\\
&& u_t = u_{xx} -\lambda u_x(x+\ln|u_x|), \quad \lambda \not =0. \label{5.4*}
\end{eqnarray}

Next, the algebra\ $A_{3.3} \oplus A_1$\ has one realization which
is the symmetry algebra of PDE belonging to the class (\ref{1.2})
$$
\langle \p_t, t \p_t +\frac{1}{2} x \p_x, -t^2 \p_t -tx
\p_x +x^2 \p_u \rangle\oplus \langle \p_u \rangle.
$$
What is more, the corresponding invariant equation
reads as
\be
\label{5.5*}
u_t = u_{xx} +\frac{1}{4} u^2_x -x^{-1} u_x +\lambda x^{-2} ,\quad
\lambda \in {\bf R}.
\ee

At last, there exists a realization of the algebra\ $A_{3.9} \oplus A_1$\
such that it is admitted by an equation of the form (\ref{1.2}), namely,
$$
\langle \p_t, \p_x, t \p_t+\frac{1}{2} x \p_x
\rangle \oplus \langle u\p_u \rangle.
$$
The corresponding invariant equation (\ref{1.2}) is given below
\be
\label{5.6*}
u_t = u_{xx} +\lambda u^{-1} u^2_x, \quad \lambda \not =0.
\ee

All other decomposable four-dimensional algebras either have no
new realizations or these realizations are not admitted by PDEs of
the form (\ref{1.2}).

Next, we carry out the complete group classification of
PDEs (\ref{5.1*})--(\ref{5.5*}).
\vspace{2mm}

\noindent
\underline{Equation (\ref{5.1*}).}

As the equation under study contains no arbitrary functions,
computing its maximal invariance algebra is an easy task.
Performing the necessary calculations in order to solve (\ref{3.6})
yields that this algebra is infinite-dimensional. The forms
of its bases operators depend essentially on $\lambda$ and are
given below
\begin{enumerate}
\item \underline{$ \lambda \not = 0, 2$}
\begin{eqnarray*}
X_1&=& -t \p_t -\frac{1}{2} x \p_x, \quad X_2 = \p_t, \quad
X_3 = \p_u, \\
X_4&=& 2 t^2 \p_t +2tx \p_x +\left[\frac{1}{2} x^2 +
(1+\lambda)t\right] \p_u, \\
X_{\infty} &=& g(t,x) e^u \p_u, \quad g_t = g_{xx}
+\frac{\lambda}{x} g_x;
\end{eqnarray*}
\item \underline{$\lambda =0.$}
\begin{eqnarray*}
X_1&=& -t \p_t -\frac{1}{2} x \p_x, \quad X_2 = \p_t, \quad
X_3 = \p_u, \\
X_4&=& 2 t^2 \p_t +2tx \p_x +\left[\frac{1}{2} x^2 +t\right] \p_u, \\
X_5&=& t \p_x +\frac{1}{2} x \p_u, \quad X_6= \p_x, \\
X_{\infty} &=& g(t,x) e^u \p_u, \quad g_t = g_{xx};
\end{eqnarray*}
\item \underline{$\lambda =2.$}
\begin{eqnarray*}
X_1&=& -t \p_t -\frac{1}{2} x \p_x, \quad X_2 = \p_t, \quad
X_3 = \p_u, \\
X_4&=& 2 t^2 \p_t +2tx \p_x +\frac{1}{2} (x^2 +3t) \p_u, \\
X_5&=& t \p_x +\frac{1}{2}\left( x +\frac{2}{x}t\right) \p_u, \quad X_6=
\p_x+\frac{1}{x} \p_u, \\
X_{\infty} &=& g(t,x) e^u \p_u, \quad g_t = g_{xx}+\frac{2}{x} g_x.
\end{eqnarray*}
\end{enumerate}

Note that the operators\ $X_1, X_2, X_3$\ and\ $X_{\infty}$\ with $g =1$
form a basis of the algebra\ $2A^1_{2.2}$.

The change of variables
$$
\bar t = t, \quad \bar x = x, \quad \bar u = u-\ln|x|,
$$
reduces the third case to the second, which means that we
have two inequivalent equations
\begin{eqnarray*}
u_t&=& u_{xx} -u^2_x; \\
u_t&=& u_{xx} +\frac{\lambda}{2} u_x -u^2_x, \quad \lambda \not=0,2.
\end{eqnarray*}
These equations are reduced to linear PDEs
\begin{eqnarray*}
v_t&=& v_{xx}; \\
v_t&=& v_{xx} +\frac{\lambda}{x} v_x, \quad \lambda \not=0, 2.
\end{eqnarray*}
with the help of the change of variables
\be
\label{5.8*}
u=-\ln |v|, \quad u=u(t,x), \quad v = v(t,x).
\ee

Thus nonlinearity in equation (\ref{5.1*}) is not essential.
\vspace{2mm}

\noindent
\underline{Equation (\ref{5.2*})}

This equation is also linearized with the aid of the change
of variables (\ref{5.8*}) to become
$$
v_t = v_{xx} +\frac{\lambda}{\sqrt{|t|}} v_x.
$$
\underline{Equation (\ref{5.3*})}

The algebra\ $2 A^3_{2.2}$ is the maximal in Lie's sense algebra
admitted by this PDE.
\vspace{2mm}

\noindent
\underline{Equation (\ref{5.4*})}

Again, the algebra\ $2 A^4_{2.2}$ is the maximal symmetry
algebra admitted by the equation in question.
\vspace{2mm}

\noindent
\underline{Equation (\ref{5.5*})}

Performing the change of variables 
$$
u = 4 \ln |v|,\quad u = u(t,x),\quad v = v(t,x),
$$
yields for the function $v$ the linear PDE
$$
v_t = v_{xx} - x^{-1} v_x +4 \lambda x^{-2} v.
$$
\underline{Equation (\ref{5.6*})}

Making the change of variables
$$
v =  \ln |u|,\quad v = v(t,x),\quad u = u(t,x),
$$
reduces the equation under study to the modified Burgers equation
$$
v_t = v_{xx} +(\lambda+1) v^2_x.
$$
The latter is locally equivalent to the linear heat conductivity equation.

Summing up, we conclude that the class of PDEs (\ref{1.2}) contains only
two equations (\ref{5.3*}) and (\ref{5.4*}) which are essentially
nonlinear and invariant under four-dimensional decomposable Lie
algebras. And what is more, the algebras\ $2 A^3_{2.2}$ and $2 A^4_{2.2}$\
are their maximal symmetry algebras.

\subsection {PDEs (\ref{1.2}) invariant under non-decomposable
four-di\-men\-si\-on\-gal Lie algebras}
\setcounter{subsection}{2}

The set of inequivalent abstract four-dimensional Lie algebras contains
ten real non-de\-com\-pos\-able Lie algebras $A_{4i} = \langle Q_1$, $Q_2,
Q_3$, $Q_4 \rangle$ $ (i=1,\ldots, 10)$ \cite{mub:}. We give below
non-zero commutation relation determining these algebras
\begin{eqnarray*}
A_{4.1}&:& [Q_2, Q_4] = Q_1, \quad [Q_3, Q_4] = Q_2; \\
A_{4.2}&:& [Q_1, Q_4] = qQ_1,\quad [Q_2, Q_4] = Q_2, \\
&& [Q_3, Q_4] = Q_2 +Q_3, \quad q\not =0;\\
A_{4.3}&:& [Q_1, Q_4] = Q_1,\quad [Q_3, Q_4] = Q_2; \\
A_{4.4}&:& [Q_1, Q_4] = Q_1,\quad [Q_2, Q_4] = Q_1+Q_2, \\
&& [Q_3, Q_4] = Q_2 +Q_3;\\
A_{4.5}&:& [Q_1, Q_4] = Q_1,\quad [Q_2, Q_4] =q Q_2, \\
&& [Q_3, Q_4] = p Q_3,\quad -1\le p \le q \le 1, \quad p q \not
=0;\\
A_{4.6}&:& [Q_1, Q_4] = qQ_1,\quad [Q_2, Q_4] = pQ_2-Q_3, \\
&& [Q_3, Q_4] = Q_2 +p Q_3,\quad q\not =0,\quad p \ge 0;\\
A_{4.7}&:& [Q_2, Q_3] = Q_1,\quad [Q_1, Q_4] = 2Q_1, \\
&& [Q_2, Q_4] = Q_2,\quad [Q_3, Q_4] = Q_2 +Q_3;\\
A_{4.8}&:& [Q_2, Q_3] = Q_1,\quad [Q_1, Q_4] = (1+q) Q_1, \\
&& [Q_2, Q_4] = Q_2 ,\quad [Q_3, Q_4] = q Q_3, \quad |q|
\le 1;\\
A_{4.9}&:& [Q_2, Q_3] = Q_1, \quad [Q_1, Q_4] = 2q Q_1, \\
&& [Q_2, Q_4] = q Q_2-Q_3,\quad [Q_3, Q_4] = Q_2+ q Q_3,
\quad q \ge 0; \\
A_{4.10}&:& [Q_1, Q_3] = Q_1,\quad [Q_2, Q_3] = Q_2, \\
&& [Q_1, Q_4] = -Q_2,\quad [Q_2, Q_4] = Q_1.
\end{eqnarray*}

Solving the above commutation relations within the class
of operators (\ref{3.5}), simplifying the obtained expressions
for\ $Q_1,\ldots, Q_4$\ with the help of appropriate equivalence
transformations and solving the invariance conditions
(\ref{3.6}) for thus obtained operators yields that there are
eleven realizations of non-decomposable four-dimensional
Lie algebras that are symmetry algebras of PDEs of the form
(\ref{1.2}). Namely,
\begin{eqnarray*}
A^1_{4.1}&=& \langle \p_u, \p_x, \p_t, t \p_x +x \p_u \rangle; \\
A^1_{4.2}&=& \langle \p_t, \p_u, \p_x, 2t \p_t +x \p_x+(u+x) \p_u
\rangle; \\
A^2_{4.2}&=& \langle \p_x, \p_u, \p_t, t \p_t +\frac{1}{2} x
\p_x+(u+t)\p_u \rangle; \\
A^1_{4.3}&=& \langle \p_u, \p_x, \p_t, t \p_t +u \p_u \rangle; \\
A^1_{4.5}&=& \langle \p_t, \p_x, \p_u, t\p_t+\frac{1}{2}x \p_x +ku \p_u
\rangle, \ k \not =0, \frac{1}{2}, 1; \\
A^1_{4.7}&=& \langle \p_u, \p_x, x\p_u-\frac{1}{2} \ln |t| \p_x, 2t
\p_t +x\p_x +2u \p_u \rangle; \\
A^1_{4.8}&=& \langle \p_x, \p_t, t\p_x + \p_u, t \p_t +\frac{1}{2} x \p_x
-\frac{1}{2} u \p_u \rangle; \\
A^2_{4.8}&=& \langle \p_u, \p_t, t\p_u+\lambda\p_x, t\p_t +
 \frac{1}{2}x \p_x +\frac{3}{2}u \p_u \rangle, \ \lambda >0; \\
A^3_{4.8}&=& \langle \p_u, \p_x, x\p_u+\lambda |t|^{\frac{1}{2}
(1-q)}\p_x, 2t \p_t +x \p_x+(1+q) u \p_u \rangle, \ |q| \not =1, \
\lambda \not =0; \\
A^4_{4.8}&=& \langle \p_u, \p_t, x\p_u+\lambda \p_t, 2t \p_t +x
\p_x+3u \p_u \rangle, \ \lambda \not =0; \\
A^1_{4.9}&=& \langle \p_u, \p_x, x\p_u+\alpha \p_x, -(\dot
\alpha)^{-1}( 1+\alpha^2) \p_t +(q-\alpha) x \p_x+[2qu -\frac{1}{2}
x^2] \p_u \rangle,
\end{eqnarray*}
where\ $q>0$ and the function\ $\alpha = \alpha(t),\ \dot\alpha \not
=0$\ is a solution of ordinary differential equation (\ref{4.51}).

Further analysis shows that PDE (\ref{1.2}) admitting the algebra\
$A_{4.1}^1$\ is linearizable. All the remaining invariant equations are
essentially nonlinear and the above algebras are their maximal in Lie's
sense symmetry algebras.

We present all the results on classification of inequivalent essentially
nonlinear PDEs (\ref{1.2}) that are invariant with respect to
four-dimensional Lie algebras (decomposable and non-decomposable) in
Table 3, where we use the following notations:
\begin{eqnarray*}
2 A^1_{2.2} &=& \langle -2t \p_t -x \p_x, \p_x, -u \p_u +\lambda
\sqrt{|t|} \p_x, \p_u \rangle, \ \lambda \not =0; \\
2 A^2_{2.2}&=& \langle \p_x -u \p_u, \p_u, \frac{1}{\lambda}
\p_t,e^{\lambda t}\p_x \rangle, \ \lambda \not =0;\\
A^1_{4.2}&=& \langle \p_t, \p_u, \p_x, 2t \p_t +x \p_x+(u+x) \p_u
\rangle; \\
A^2_{4.2}&=& \langle \p_x, \p_u, \p_t, t \p_t +\frac{1}{2} x
\p_x+(u+t)\p_u \rangle; \\
A^1_{4.3}&=& \langle \p_u, \p_x, \p_t, t \p_x +u \p_u \rangle; \\
A^1_{4.5}&=& \langle \p_t, \p_x, \p_u, t\p_t+\frac{1}{2}x \p_x +ku \p_u
\rangle, \ k \not =0, \frac{1}{2}, 1; \\
A^1_{4.7}&=& \langle \p_u, \p_x, x\p_u-\frac{1}{2} \ln |t| \p_x, 2t
\p_t +x\p_x +2u \p_u \rangle; \\
A^1_{4.8}&=& \langle \p_x, \p_t, t\p_x+ \p_u, t \p_t +\frac{1}{2} x \p_x
-\frac{1}{2} u \p_u \rangle; \\
A^2_{4.8}&=& \langle \p_u, \p_t, t\p_u+\lambda\p_x, t\p_t+\frac{1}{2}x \p_x
+\frac{3}{2}u \p_u \rangle, \ \lambda >0; \\
A^3_{4.8}&=& \langle \p_u, \p_x, x\p_u+\lambda |t|^{\frac{1}{2}
(1-q)}\p_x, 2t \p_t +x \p_x+(1+q) u \p_u \rangle, \ |q| \not =1, \
\lambda \not =0; \\
A^4_{4.8}&=& \langle \p_u, \p_t, x\p_u+\lambda \p_t, 2t \p_t +x
\p_x+3u \p_u \rangle,\ \lambda \not =0; \\
A^1_{4.9}&=& \langle \p_u, \p_x, x\p_u+\alpha \p_x, -(\dot
\alpha)^{-1}( 1+\alpha^2) \p_t +(q-\alpha) x \p_x+[2qu -\frac{1}{2}
x^2] \p_u \rangle, \\
&&\mbox{where}\ q>0\ \mbox{and}\ \alpha = \alpha(t),\ \dot
\alpha \not =0\ \mbox{is a solution of (\ref{4.51})};\\
AG^1_3(1,1)&=& \langle \p_x, t \p_x +\p_u,\p_t, -2t \p_t -x \p_x +u
\p_u, t^2 \p_t +tx \p_x-(tu-x) \p_u \rangle.
\end{eqnarray*}

\section{Further algebraic analysis}
\setcounter{section}{6}
\setcounter{equation}{0}

In this section we prove that there are no essentially nonlinear
equations of the form (\ref{1.2}) that admit invariance algebra of the
dimension higher than 4. This means that the above obtained group
classification of invariant PDEs (\ref{1.2}) is complete.

Our considerations are purely algebraic and are based on the
Levi-Maltsev theorem claiming that any Lie algebra  over the field ${\bf
R}$ or ${\bf C}$ can be decomposed into a semi-direct sum of a maximal
solvable ideal $N$ and semi-simple subalgebra $S$. This means that the
problem of classification of abstract Lie algebras reduces to
classifying
\begin{itemize}
\item{solvable Lie algebras,}
\item{semi-simple Lie algebras,}
\item{algebras that are semi-direct sums of semi-simple and solvable
Lie algebras.}
\end{itemize}

We consider the above enumerated cases separately.
\vspace{2mm}

\noindent
{\bf Case 1.}\ Solvable Lie algebras.

As far as we know, the problem of classification of abstract solvable
real  Lie algebras has been completely solved for solvable Lie algebras
of the dimension up to five (see, e.g., \cite{mub:, mub1:}). For higher
dimensional solvable Lie algebras only partial results have been
obtained \cite{mor,mub2:}. The main difficulty is that a number of
non-isomorphic solvable $n$-dimensional Lie algebras increases rapidly
with increasing $n$. For example, there are 67 types of five-dimensional
solvable Lie algebras \cite{mub1:} and 99 types of six-dimensional
solvable Lie algebras having a nilpotent element \cite{mub2:}. This is
why, the problem of exhaustive classification of solvable Lie algebras
of the dimension $n>5$ is a \lq wild problem\rq . However, in the case
under study it is possible to carry out such a classification due to the
fact that we are looking for rather specific realizations of the
solvable Lie algebras.

\newpage
\begin{center}
{\it Table 3.}\ {\bf Nonlinear PDEs (\ref{1.2}) admitting
four-dimensional Lie algebras}
\end{center}
\begin{center}
\begin{tabular}{| l | c| c|}   \hline
& &\\
No. & Equation & Maximal \\  %[2mm]
& & invariance  \\ % [2mm]
& & algebra \\  [2mm] \hline
& &\\
1& $u_t = u_{xx} +\frac{\lambda \epsilon u_x}{4 \sqrt{|t|}} \ln |t
u^2_x| +\frac{\beta u_x}{\sqrt{|t|}},$&$2 A^1_{2.2}$ \\ [2mm]
& & \\
& $\epsilon =1$ for $t>0$, \ $\epsilon =-1$ for $t<0$, \ $\beta \in
R, \ \lambda \not = 0$& \\ [2mm] \hline
& &\\
2& $u_t = u_{xx} -\lambda u_x(x+\ln|u_x|), \ \lambda \not =0$&$2
A^2_{2.2}$ \\ [2mm] \hline
& &\\
3& $u_t = u_{xx} +\lambda \exp(-u_x), \ \lambda \not =0$&$
A^1_{4.2}$ \\ [2mm] \hline
& &\\
4& $u_t = u_{xx} +2\ln|u_x|$&$ A^2_{4.2}$ \\ [2mm] \hline
& &\\
5& $u_t = u_{xx} - u_x\ln|u_x|+\lambda u_x, \ \lambda \in {\bf R}$&$
A^1_{4.3}$ \\ [2mm] \hline
& &\\
6& $u_t = u_{xx} +\lambda u_x^{\frac{2k-2}{2k-1}},\ \lambda \not =0,
\ k \not =0, \frac{1}{2}, 1$&$ A^1_{4.5}$ \\ [2mm] \hline
& &\\
7& $u_t = u_{xx} +\frac{1}{4t} u^2_x$&$
A^1_{4.7}$ \\ [2mm] \hline
& &\\
8& $u_t = u_{xx} -u u_x+\lambda |u_x|^{\frac{3}{2}}$& \\ [2mm]
& & \\
&$ \lambda \not =0$&$A^1_{4.8}$\\ [2mm]
& & \\
& $\lambda  =0$&$AG^1_3(1,1)$\\ [2mm] \hline
9& $u_t = u_{xx} +\lambda^{-1} x +m \sqrt{|u_x|}, \ \lambda >0, \ m
\not =0$&$A^2_{4.8}$ \\ [2mm] \hline
\end{tabular}
\end{center}

\begin{center}
\begin{tabular}{| l | c| c|}   \hline
& &\\
No. & Equation & Maximal \\  %[2mm]
& & invariance  \\ % [2mm]
& & algebra \\  [2mm] \hline
10& $u_t = u_{xx} -\frac{\lambda \epsilon}{4}(1-q)
|t|^{-\frac{1}{2}(1+q)}u^2_x$ & \\[2mm]
&& \\
&$\lambda \not =0, \ |q| \not =1, \ \epsilon =1$ для $t>0, \ \epsilon
=-1$ для $t<0$&$A^3_{4.8}$ \\ [2mm] \hline
&&\\
11& $u_t = u_{xx} +m \sqrt{|t-\lambda u_x|}, \ \lambda \cdot m \not
=0$&$ A^4_{4.8}$ \\ [2mm] \hline
&&\\
12& $u_t = u_{xx} -\frac{1}{2} \dot \alpha u^2_x +(\lambda -\alpha)
(1+\alpha^2)^{-1}, \ \lambda \in {\bf R}$&$A^1_{4.9}$ \\ [2mm] \hline
\end{tabular}
\end{center}
\vspace{2mm}

Our considerations are based on the well-known fact that for any
solvable Lie algebra $L_n$ with dim\, $L_n = n$ over the field of real
numbers we can construct a composition series for $L_n$
\[
L_0\subset L_1\subset \cdots \subset L_{n-1} \subset L_n,
\]
where each algebra\ $L_i,\ {\rm dim}\,L_i=i\ (1=0,1,\ldots, n-1)$\ is an
ideal in the algebra $L_{i+1}$. Hence we easily get the following
assertion. Suppose that there exist realizations\ $A^1, A^2,\ldots ,
A^{N}$\ of solvable Lie algebras within a given class of Lie vector
fields ${\cal V}$ of the dimension not greater than $m$ and,
furthermore, realizations of the dimension $m+1$ do not exist. Then the
realizations\ $A^1, A^2,\ldots , A ^{N}$\ exhaust a set of all possible
realizations of solvable Lie algebras within the class ${\cal V}$.

According to the results of Section 5 there are twelve realizations
of solvable four-dimensional Lie algebras within the class of operators
(\ref{3.5}). If we will prove that there are no realizations of solvable
five-dimensional Lie algebras within the class (\ref{3.5}) which are
invariance algebras of PDE of the form (\ref{1.2}), then in view of
the above assertion we conclude that the obtained realizations
of solvable Lie algebras of the dimension $n\le 4$ exhaust the
set of all possible realizations of solvable Lie algebras in the
case under study.

First, we investigate the case when a five-dimensional solvable Lie
algebra is a direct sum of four- and one-dimensional solvable Lie
algebras. We consider in more detail the realization $2A^1_{2.2}$, where
\[
e_1=-2t\p_t - x\p_x,\quad e_2=\p_x,\quad e_3=-u\p_u +
\lambda\sqrt{|t|}\p_x,\quad e_4=\p_u.
\]
Taking the basis element $e_5$ in the general form (\ref{3.5})
\[
e_5=2a(t)\p_t + (\dot a(t) x +b(t))\p_x + f(t,x,u)\p_u
\]
and checking the commutation relations\ $[e_i, e_5]=0,\ (i=1,2,3,4)$\
yield that\ $e_5=\sqrt{|t|}\p_x$. Inserting this expression into
the invariance criterion (\ref{3.6}) we arrive at the contradictory
equality
\[
\frac{1}{2\sqrt{|t|}}u_x=0.
\]
Consequently, the algebra\ $A_{2.2}^1$\ cannot be extended to a
realization of five-dimensional solvable Lie algebra admitted by PDE of
the form (\ref{1.2}). The same assertion holds for the realizations\
$2A^2_{2.2}, A^1_{4.7}, A^3_{4.8}, A^4_{4.8}, A^1_{4.9}$.

Furthermore, the realizations $A^1_{4.2}, A^2_{4.2}, A^1_{4.3},
A_{4.5}^1, A^1_{4.8}, A^2_{4.8}$ cannot be extended to realizations of
five-dimensional solvable Lie algebras within the class of operators
(\ref{3.5}).

Next, we turn to the case of indecomposable five-dimensional solvable
Lie algebras. According to the classification given in \cite{mub1:}
there are five types of indecomposable five-dimensional solvable Lie
algebras
\begin{enumerate}
\item[{1)}]{nilpotent algebras,}
\item[{2)}]{algebras having one non-nilpotent basis element and containing
the commuting ideal\ $4A_1$,}
\item[{3)}]{algebras having one non-nilpotent basis element and containing
the ideal\ $A_{3.5}\oplus A_1$,}
\item[{4)}]{algebras having one non-nilpotent basis element and containing
the ideal\ $A_{4.1}$,}
\item[{5)}]{algebras having two nil-independent basis elements (two basis
elements are called nil-independent if there is no linear combination
of these which is nilpotent).}
\end{enumerate}

Five-dimensional solvable algebras of the first type contain either
a four-dimensional commuting radical or a radical that is isomorphic to
the decomposable algebra\ $A_1\oplus A_{3.5}$. Consequently, realizations
of these algebras which could be invariance algebras of PDE of the form
(\ref{1.2}) do not exist. Similar reasonings yield the same
statement for the algebras of the second, third and fourth types.

Consider the algebras of the fifth type. Let\ $e_1, e_2, e_3, e_4, e_5$\
form a basis of an algebra of this type. Then inequivalent abstract
five-dimensional solvable Lie algebras having two nil-independent basis
elements read
\begin{eqnarray*}
L_1&:& [e_1, e_4]=e_1,\quad [e_3, e_4]=\beta e_3,\quad [e_2, e_5]=e_2,\\
&&[e_3, e_5]=\gamma e_3,\quad \beta^2+\gamma^2\ne 0;\\
L_2&:& [e_1, e_4]=\alpha e_1,\quad [e_2, e_4]= e_2,\quad [e_3, e_4]=e_3,\\
&&[e_1, e_5]= e_1,\quad [e_3, e_5]=e_2;\\
L_3&:& [e_1, e_4]=\alpha e_1,\quad [e_2, e_4]= e_2,\quad [e_3, e_4]=e_3,\\
&&[e_1, e_5]=\delta e_1,\quad [e_2, e_5]=-e_3,\quad [e_3, e_5] = e_2,\quad
\alpha^2+\delta^2\ne 0;\\
L_4&:& [e_2, e_3] = e_1,\quad [e_1, e_4]= e_1,\quad [e_2, e_4]=e_2,\\
&&[e_2, e_5]=- e_2,\quad [e_3, e_5]=e_3;\\
L_5&:& [e_2, e_3]=e_1,\quad [e_1, e_4]= 2e_1,\quad [e_2, e_4]=e_2,\\
&&[e_3, e_4]= e_3,\quad [e_2, e_5] = - e_3,\quad [e_3, e_5] = e_2;\\
L_6&:& [e_1, e_4]=e_1,\quad [e_2, e_5]= e_2,\quad [e_4, e_5]=e_3;\\
L_7&:& [e_1, e_4]=e_1,\quad [e_2, e_4]= e_2,\quad [e_1, e_5]= -e_2,\\
&&[e_2, e_5]= e_1,\quad [e_4, e_5]=e_3.
\end{eqnarray*}
Note that we give non-zero commutation relations only.

The algebra $L_1$ contains a radical isomorphic to the decomposable
four-dimensional Lie algebra\ $A_1\oplus A_{3.9}$. Next, the algebra
$L_6$ contains a radical isomorphic to the algebra\ $2A_1\oplus A_{2.2}$.
At last, the algebra $L_7$ contains a radical isomorphic to the algebra\
$A_1\oplus A_{3.7}$. Hence we conclude that there are no realizations
of the algebras\ $L_1, L_6, L_7$\ which are invariance algebras of
PDE (\ref{1.2}).

The algebra $A^1_{4.5}$ gives a realization of the algebra $A_{4.5}$
with\ $q=\frac{1}{2}, p\ne 0, \frac{1}{2}, 1$. On the other hand, the
algebras $L_2, L_3$ contain a radical isomorphic to the algebra
$A_{4.5}$ with $q=1$. Hence it follows that the realization $A^1_{4.5}$
cannot be extended to yield a realization of the algebras $L_2, L_3$.

The algebra $L_4$ contains a radical isomorphic to the algebra\ $A_{4.8}$\
with $q=0$. To meet this requirement we have to choose for this radical
the realization\ $A^3_{4.8}$\ with $q=0$, namely,
\[
e_1=\p_u,\quad e_2=\p_x,\quad e_3=x\p_x + \lambda |t|^{\frac{1}{2}}\p_x,
\quad e_4=2t\p_t + x\p_x + u\p_u.
\]
Checking commutation relations for an operator $e_5$ of the form (\ref{3.5})
shows that the realization\ $A^3_{4.5}$\ cannot be extended to give a
realization of the five-dimensional algebra $L_4$.

The algebra $L_5$ contains a radical isomorphic to the algebra\
$A_{4.8}\ (q=1)$.\ However, there are no realizations of the algebra\
$A_{4.8}$\ which might yield a realization of this radical. Hence we
conclude that there are no realizations of the algebra $L_5$ within
the class of operators (\ref{3.5}).
\vspace{2mm}

\noindent
{\bf Case 2.}\ Semi-simple Lie algebras.

As proved by Cartan, any real or complex semi-simple Lie algebra is
decomposed into a direct sum of mutually orthogonal simple algebras. In
view of this fact, the problem of classification of abstract semi-simple
Lie algebras reduces to classifying simple Lie algebras (see, e.g.
\cite{bar}). The classification of simple Lie algebras is well-known.
There are four series of non-exceptional complex simple Lie algebras
$A_n,\ B_n,\ C_n,\ D_n$ and five types of exceptional Lie algebras.

The lower dimensional semi-simple Lie algebras are connected by
the following isomorphisms \cite{bar}:
\begin{eqnarray}
&&su(2) \sim so(3)\sim sp(1),\quad sl(2,{\bf R})\sim su(1,1)\sim
so(2,1)\sim sp(1, {\bf R}),\nonumber\\
&&so(5)\sim sp(2),\quad so(3,2)\sim sp(2,{\bf R}),\quad
so(4,1)\sim sp(1,2),\nonumber\\
&&so(4)\sim so(3)\oplus so(3),\quad so(2,2)\sim sl(2,{\bf R})
\oplus sl(2,{\bf R}),\label{9.1}\\
&&sl(2,{\bf C})\sim so(3,1),\quad su(4)\sim so(6),\quad
sl(4,{\bf R})\sim so(3,3),\nonumber\\
&&su(2,2)\sim so(4,2),\quad su(3,1)\sim so^*(6),\quad su^*(4)\sim
so(5,1).\nonumber
\end{eqnarray}

It turns out that\ $sl(2,{\bf R})\sim su(1,1)\sim so(2,1)\sim sp(1, {\bf
R})$ are the only real forms of the algebras given in (\ref{9.1}) that
have realizations within the class of operators (\ref{3.5}). The reason
is that all other algebras contain the subalgebra $so(3)$ and the latter
has no realizations within the class (\ref{3.5}). Next, all the real
forms of higher dimensional non-exceptional simple Lie algebras contain
the algebra $so(3)$ as a subalgebra. Consequently, they have no
realizations within the class of operators (\ref{3.5}).

The exceptional simple Lie algebras have no realizations within the class
of differential operators of the form (\ref{3.5}).

Consequently, the only semi-simple algebras that might be admitted
by PDE of the form (\ref{1.2}) are algebras of the form
\[
sl(2, {\bf R}),\quad sl(2, {\bf R})\oplus sl(2, {\bf R}),\quad
sl(2, {\bf R})\oplus sl(2, {\bf R})\oplus sl(2, {\bf R}),\ldots
\]
As straightforward calculation shows, there are no PDEs of
the form (\ref{1.2}) invariant with respect to the algebra\ $sl(2, {\bf R})
\oplus sl(2, {\bf R})$. Hence it follows, that the only semi-simple
algebra that might be admitted by (\ref{1.2}) is the three-dimensional
algebra\ $sl(2, {\bf R})$.
\vspace{2mm}

\noindent
{\bf Case 3}. Semi-direct sums of semi-simple and solvable algebras.

The algebras of the type considered can be split into two classes.
\begin{itemize}
\item{ algebras which are decomposable into direct sums of semi-simple
and solvable algebras,}
\item{algebras which cannot be decomposed into direct sums of semi-simple
and solvable algebras.}
\end{itemize}

As shown above, there exists only one realization\ $A^1_{3.3}$ of a
 semi-simple algebra which is an invariance algebra of an equation of
the form (\ref{1.2}). It is a realization of simple algebra $A_{3.3}$
isomorphic to the algebra $sl(2, {\bf R})$. If we will try to extend
this realization to get a realization of a direct sum of semi-simple and
solvable Lie algebras, then we will have to stop at the first step,
since the realization\ $A^1_{3.3}\oplus A_1$\ is an invariance algebra
of linear PDE (see Section 5).

Turn now to the algebras which are not decomposable into a direct sum of
semi-simple and solvable Lie algebras. According to the above results of
the previous two cases, their dimension cannot be higher than $3+4=7$.
In the paper \cite{tur:} a complete classification of the algebras which
are semi-direct sums of semi-simple and solvable Lie algebras and have
the dimension $n\le 8$ is obtained. Analysis of these algebras shows
that they have no realizations within the class of operators (\ref{3.5})
that are invariance algebras of PDE of the form (\ref{1.2}).

Summing up we conclude that there are no real Lie algebras of the
dimension $n\ge 5$ which are invariance algebras of essentially
nonlinear PDEs belonging to the class (\ref{1.2}). This means that our
classification of nonlinear PDEs (\ref{1.2}) invariant under the one-,
two-, three- and four-dimensional Lie algebras gives the complete
description of heat equations (\ref{1.2}) possessing non-trivial Lie
symmetries.

\section{Comparison to other classifications}
\setcounter{section}{7}
\setcounter{equation}{0}

Here we briefly review the earlier results on classification of
invariant PDEs belonging to the class (\ref{1.2}). We will show that all
of them can be derived from equations given in Tables 1--3 (either
directly or via local transformations of dependent and independent
variables).

The problem of group classification of the nonlinear heat conductivity
equation with a nonlinear convection term
\be
\label{8.1}
u_t = [K(u) u_x]_x +[\Phi(u)]_x
\ee
has been considered in \cite{or,ed}. Evidently, provided\ $K(u)=1$,\
it is included into the class (\ref{1.2}).

Next, Dorodnitsyn \cite{dor} has classified invariant nonlinear
heat conductivity equations with nonlinear source
\be
\label{8.2}
\frac{\p T}{\p t}= \frac{\p}{\p x} \left(K(T) \frac{\p T}{\p x}\right)
+Q(T).
\ee
Again, this equation with\ $K(u)=1$\ belongs to the class (\ref{1.2}).
Note that an analogous problem for the two- and three-dimensional
PDEs of the type (\ref{8.2}) has been solved in \cite{dor1}.

The papers \cite{ser} are devoted to symmetry analysis of nonlinear
PDEs of the form
\be
\label{8.3}
u_t = [A(u) u_x]_x +B(u) u_x +C(u).
\ee

Nonlinear PDE (\ref{8.3}) is a natural generalization of equations
(\ref{8.1}), (\ref{8.2}) and, furthermore, is contained in the class of
PDEs (\ref{1.2}) provided\ $A(u)=1$.

Gandarias \cite{gan} has carried out group classification of
equation
\be
\label{8.4}
u_t =(u^n)_{xx}+g(x) u^m+f(x) u^s u_x, \quad n\not =0
\ee
that is also included into the class (\ref{1.2}), provided
the condition\ $n=1$ holds.

\subsection{Group analysis of equation (\ref{8.1})}
\setcounter{subsection}{1}

According to \cite{or,ed} the results on group classification of
equation (\ref{8.1}) under\ $K(u)=1$,\ namely of equation
\be
\label{8.5}
u_t = u_{xx} +[\Phi(u)]_x,
\ee
can be summarized as follows. The maximal invariance algebra
admitted by PDE (\ref{8.5}) under an arbitrary function\
$\Phi,\ \frac{d \Phi}{du} \not =0$\ is the two-dimensional Lie algebra\
$\langle \p_t, \p_x\rangle $. Extension of the invariance
algebra is only possible, provided
\begin{eqnarray*}
&1)& \Phi = \beta u^\nu, \quad Q_{\rm new}=2(1-\nu) t \p_t +(1-\nu) x
\p_x +u \p_u;\\
&2)& \Phi =\beta \ln u, \quad Q_{\rm new}=2t \p_t+x\p_x +u\p_u; \\
&3)& \Phi =\beta e^{\nu u}, \quad Q_{\rm new}=t \p_t +\frac{1}{2} x
\p_x -\frac{1}{2 \nu} \p_u.
\end{eqnarray*}
Here\ $\nu\not =0,1,2$,\ $\beta \in {\bf R}$.

Note that equation (\ref{8.5}) with\ $\Phi = \beta u^\nu$,\ where
$\nu=2$ coincides with the Burgers equation which maximal symmetry
algebra is five-dimensional and is isomorphic to the full Galilei
algebra.

For the first case\ $(\Phi = \beta u^\nu)$\ the invariance algebra
is isomorphic to the algebra\ $A_{3.9}$\ $(q=\frac{1}{2})$. This
isomorphism is established by choosing the basis operators
as follows
$$
Q_1=\p_t,\quad Q_2=\p_x,\quad
Q_3=t \p_t+\frac{1}{2} x\p_x +\frac{1}{2(1-\nu)} u
\p_u.
$$
Furthermore the change of variables
$$
t = t, \quad x=x, \quad v = u^{2(1-\nu)}
$$
transforms the above realization to become\ $A^2_{3.9}$.\ With
this transformation the corresponding invariant equation (\ref{8.5})
takes the form
\be
\label{8.6}
v_t = v_{xx} +\frac{2\nu-1}{2(1-\nu)} v^{-1} v^2_x +\beta \frac{2
\nu}{2(1-\nu)} v^{-\frac{1}{2}} v_x,
\ee
which is a particular case of the equation invariant with
respect to the algebra\ $A^2_{3.9}$\ from Table 2.

Given the condition\ $\Phi = \beta \ln u$, the invariance algebra
of (\ref{8.5}) is also isomorphic to the algebra\ $A_{3.9}$\
($q=\frac{1}{2}$), its basis being chosen in the following
way:
$$
\p_t,\quad \p_x,\quad t \p_t+\frac{1}{2} x \p_x +\frac{1}{2} u \p_u.
$$
The change of variables
$$
t = t,\quad x=x,\quad v = u^2
$$
reduces the corresponding invariant equation (\ref{8.5}) to the
form
$$
v_t = v_{xx} -\frac{1}{2} v^{-1} v^2_x +\beta v^{-\frac{1}{2}} v_x.
$$
The latter is, evidently, a particular case of PDE invariant
with respect to the algebra\ $A^2_{3.9}$\ from Table 2.

At last, for the third case the invariance algebra is also isomorphic to
the algebra\ $A_{3.9}\ (q=\frac{1}{2})$ and is reduced to the
realization\ $A^2_{3.9}$\ with the help of the change of variables\ $t =
t, \ x=x,\ v=e^{-2\nu u}$.\ The corresponding invariant equation
(\ref{8.5}) with this change of variables takes the form
$$
v_t = v_{xx} -v^{-1} v^2_x +\beta \nu v^{-\frac{1}{2}} v_x,
$$
which is a particular case of PDE invariant with respect to the algebra\
$A^2_{3.9}$\ from Table 2.

Summing up we conclude that the group classification
of PDE (\ref{8.5}) within the equivalence relation follows
from our classification of equations invariant under the Lie algebra\
$A^2_{3.9}$ if we put in these\
$$
G(\om) = \lambda_1 \om +\lambda_2 \sqrt{\om},\quad
\om=u^{-1} u^2_x, \quad \{\lambda_1, \lambda_2\} \subset {\bf R}.
$$

\subsection{Group analysis of equation (\ref{8.2})}

The results on group classification of (\ref{8.2}) with\ $K(T)=1$,
namely for PDE  of the form
\be \label{8.7}
u_t = u_{xx} +F(u)
\ee
given in \cite{dor} can be formulated in the following way.
Provided the function\ $F,\ \frac{d^2 F}{d u^2} \not =0$\
is arbitrary, the maximal invariance algebra of (\ref{8.7})
is the two-dimensional Lie algebra\ $\langle \p_t, \p_x
\rangle$. Extension of the invariance algebra is only
possible provided
\begin{eqnarray*}
&1)& F = \pm e^u,\quad Q_{\rm new}=t \p_t +\frac{1}{2}
     x \p_x -\p_u; \\
&2)& F = \pm u^n,\quad Q_{\rm new}=t \p_t +\frac{1}{2} x \p_x
-\frac{1}{n-1} u \p_u; \\
&3)& F = \delta u \ln u,\quad \delta = \pm 1,\quad
     Q^1_{\rm new}=e^{\delta t}[\p_x -\frac{\delta}{2}
     x u \p_u],\quad Q^2_{\rm new}=e^{\delta t} u \p_u.
\end{eqnarray*}
Note that the classification results yielding linear
invariant PDEs are neglected here.

Consider first the case 1. Then the change of variables
$$
t = t,\quad x=x,\quad v=e^{-u}
$$
reduces the invariance algebra to become\ $A^2_{3.5}$\
and, furthermore, the corresponding invariant equation (\ref{8.7})
takes the form 
$$
v_t = v_{xx} -v^{-1} v^2_x \mp 1.
$$

For the second case, there is the change of variables
$$
t = t,\quad x=x,\quad v=u^{1-n},\quad n \not= 1
$$
that reduces the invariance algebra to become\ $A^2_{3.5}$.
The corresponding invariant equation (\ref{8.7}) takes the form
$$
v_t = v_{xx} +\frac{n}{n-1} v^{-1} v^2_x \pm \frac{1}{1-n}.
$$

The above two PDEs are particular cases of the equation invariant with
respect to the algebra\ $A^2_{3.5}$\ from Table 2.

At last, in the third case the maximal invariance algebra is
four-di\-men\-sion\-al. Making use of the change of variables
$$
\tau = -\frac{\delta}{2} e^{-2 \delta t},\quad \xi =
e^{-\delta t}x, \quad v = e^{-\delta t}[ \ln |u|
+\frac{\delta}{4} x^2]
$$
we become convinced of the fact that the invariance algebra is equivalent
to\ $A^3_{4.8}$ with\ $q=0,\ \lambda =2 \delta \sqrt{2}$.\ The
corresponding invariant equation (\ref{8.7}) is reduced to the form
$$
v_\tau = v_{\xi \xi} -\frac{\varepsilon \delta}{2}
|\tau|^{-\frac{1}{2}} (v_\xi)^2 +\frac{\varepsilon \delta}{2 \sqrt{2}}
|\tau|^{-\frac{1}{2}},
$$
where $\varepsilon =1$ for $\tau>0$ and $\varepsilon =-1$ for $\tau<0$.
Making the second change of variables
$$
\tau = \tau,\quad \xi = \xi,\quad \om =
v+\frac{\delta}{\sqrt{2}} |\tau|^{\frac{1}{2}}
$$
yields the equation under the number 9 from Table 3 with\
$\lambda = 2\sqrt{2} \delta,\ q=0$ and
$$
\om_\tau = \om_{\xi \xi}-\frac{\varepsilon \delta}{\sqrt{2}}
|\tau|^{-\frac{1}{2}} (\om_\xi)^2.
$$

Similar analysis of classification results for PDEs (\ref{8.3})
\cite{ser} and (\ref{8.4}) \cite{gan} shows that all the invariant
equations obtained there can be derived from invariant PDEs given in
Tables 2, 3 under appropriate changes of variables. We unable to present
here the corresponding calculations in a compact form, since they are
extremely lengthy (just a precise formulation of classification results
obtained in \cite{ser,gan} requires several pages, to say nothing of a
space needed to give a detailed analysis of these).

\section{Concluding Remarks}

We have carried out group classification of nonlinear heat transfer
equations of the form (\ref{1.2}) and proved that essentially nonlinear
PDEs (\ref{1.2}) admit at most four-parameter invariance group.
Furthermore, we have established that there are three classes of
equations (\ref{1.2}) invariant with respect to one-parameter groups
(formulae (\ref{4.4})--(\ref{4.6})), seven classes of equations
(\ref{1.2}) invariant with respect to two-parameter groups (formulae
(\ref{5.1})--(\ref{5.3}), (\ref{5.5}), (\ref{5.6}), (\ref{5.8}),
(\ref{5.9})), twenty eight classes of equations (\ref{1.2}) invariant
with respect to three-parameter groups (Tables 1, 2) and twelve classes
of equations (\ref{1.2}) invariant with respect to four-parameter groups
(Table 3).

We concentrate on studying essentially nonlinear heat conductivity
equations since the linear case is well investigated. However, it is
fairly simple to recover the corresponding results within the framework
of our approach. Consider the most general linear PDE of the parabolic type
in one spatial variable
\begin{equation}
\label{c1}
u_t=f(t,x)u_{xx} + g(t,x)u_x + h(t,x)u.
\end{equation}

The most general infinitesimal operator of the symmetry group admitted
by (\ref{c1}) reads as
\[
Q=T(t)\p_t+X(t,x)\p_x+(U(t,x)u+u_0(t,x))\p_u,
\]
where\ $T, X, U$\ are arbitrary smooth functions and $u_0$ is an
arbitrary solution of (\ref{c1}). As usual, we neglect the trivial
symmetry\ $u_0(t,x)\p_u$\ and put\ $u_0=0$. Next, the equivalence group
of the class of PDEs (\ref{c1}) has the form
\[
\bar t=F(t),\quad \bar x=G(t,x),\quad \bar u=H(t,x)u.
\]
Using these facts it is straightforward to check that the list of
inequivalent one-dimensional Lie algebras admitted by (\ref{c1}) is
exhausted by the following three algebras
\[
A_1=\langle \p_x\rangle,\quad A_2=\langle \p_t\rangle,\quad
A_3=\langle U(t,x)\p_u\rangle.
\]

As equation (\ref{c1}) is linear, it admits the one-dimensional
Lie algebra $u\p_u$ with arbitrary $f, g, h$. Consequently, any
two-dimensional algebra is reduced to the one of three possible
inequivalent forms\ $\langle u\p_u\rangle \oplus A_i\ (i=1,2,3)$.

If equation (\ref{c1}) is invariant with respect to the algebra\
$\langle u\p_u\rangle \oplus A_1$,\ then its coefficients are
independent of $x$. Hence we easily get that it is reduced to
the standard heat transfer equation
\begin{equation}
\label{c2}
u_t=u_{xx}.
\end{equation}

Turn next to the case of the algebra $\langle u\p_u\rangle \oplus A_2$.
Now the coefficients of (\ref{c1}) are independent of $t$ and,
therefore, this equation can be reduced to become
\begin{equation}
\label{c3}
u_t=u_{xx}+V(x)u
\end{equation}
with an arbitrary smooth function $V$. It is a common knowledge
that the above PDE has a symmetry algebra of the dimension higher than
$2$ if and only if
\begin{equation}
\label{c4}
V(x)=\frac{\lambda_0}{x^2} + \lambda_1x^2 +\lambda_2 x + \lambda_3,
\end{equation}
where $\lambda_0,\ldots, \lambda_3$ are arbitrary constants with
$\lambda_0\lambda_2=0$. Furthermore, provided $\lambda_0=0$, PDE
(\ref{c3}), (\ref{c4}) is equivalent to the heat transfer equation
(\ref{c2}). If, $\lambda_0\ne 0$, then PDE (\ref{c3}), (\ref{c4})
reduces to the following equation:
\begin{equation}
\label{c5}
u_t=u_{xx} + \frac{\lambda_0}{x^2}u
\end{equation}
which is invariant under the four-dimensional Lie algebra
\[
\left\langle \p_t,\ 2t\p_t + x\p_x,\ t^2\p_t + tx\p_x -
\left(\frac{t}{2} +\frac{x^2}{4}\right)u\p_u, u\p_u\right\rangle.
\]

Summing up we conclude that there are three inequivalent classes of PDEs
(\ref{c1}) whose symmetry algebras have the dimensions higher than one,
namely, the heat transfer equation (\ref{c2}) admitting the
six-dimensional Lie algebra, equation (\ref{c5}) invariant with respect
to the  four-dimensional algebra and equation (\ref{c3}) that admits the
two-dimensional algebra\ $\langle\p_t, u\p_u\rangle$. This completes
group classification of heat transfer equations (\ref{1.2}) admitting
nontrivial Lie symmetry.

When classifying invariant equations (\ref{1.2}) we utilize as
equivalence transformations local transformations of dependent and
independent variables. Using non-local transformations, on the one hand,
may result in reduction of equivalence classes and, on the other hand,
may yield so-called quasi-local symmetries (for more detail on on
quasi-local symmetries see, e.g. \cite{ah1:}). Consider, as an example,
the following subclass of PDEs of the form (\ref{1.2}):
\begin{equation}
\label{c6}
u_t=u_{xx} + f_1(t)u+f_2(t,x,u_x)
\end{equation}
with arbitrary smooth functions $f_1, f_2$. If we differentiate
(\ref{1.2}) with respect to $x$ and make a change of the dependent
variable
\begin{equation}
\label{c7}
u_x(t,x)\to v(t,x),
\end{equation}
then we get a subclass of quasi-linear PDEs of the form (\ref{1.2})
\begin{equation}
\label{c8}
v_t=v_{xx} + f_1(t)v + f_{2x}(t,x,v) + f_{2v}(t,x,v)v_x. 
\end{equation}
Evidently, the above two classes of PDEs (\ref{c6}) and (\ref{c7})
are inequivalent in the sense of the definition given in Section 3,
since transformation (\ref{c7}) is not local. 

The technique developed in the present paper can be efficiently applied
to carry out group classification of arbitrary classes of PDEs in two
independent variables, since their maximal symmetry algebras are, as a
rule, low dimensional and we can use the classification of abstract low
dimensional Lie algebras.

These and the related problems are under study now and the results will
be reported in our future publications.

\end{document}